\documentclass[11pt,twoside]{book}
\usepackage{asp2010}
\def\mathbi#1{\textbf{\em #1}}

\def\spose#1{\hbox to 0pt{#1\hss}}
\def\simlt{\mathrel{\spose{\lower 3pt\hbox{$\mathchar"218$}}
     \raise 2.0pt\hbox{$\mathchar"13C$}}}
\def\simgt{\mathrel{\spose{\lower 3pt\hbox{$\mathchar"218$}}
     \raise 2.0pt\hbox{$\mathchar"13E$}}}
\def\lsim{\rlap{$<$}{\lower 1.0ex\hbox{$\sim$}}}
\def\gsim{\rlap{$>$}{\lower 1.0ex\hbox{$\sim$}}}

\begin{document}

\title{Heliophysics gleaned from seismology}

\author{D.O. Gough\\ 
Institute of Astronomy\\
\& Department of Applied Mathematics and Theoretical Physics,\\
Madingley Road, Cambridge, CB3 0HA, UK}

\date{}

\begin{abstract}
Some of the principal heliophysical\footnote{The term heliophysics was coined in 1981 to denote the physics of the entire sun, out to the corona.  It is a direct translation from the French `h\'eliophysique', which was introduced to provide a distinction from physique solaire (solar physics) which in practice was then confined to only the outer layers.  It is a subdiscipline of heliology \citep[cf.][]{jcddogheliologicalinverseproblem1976Natur}.   Recently the meaning of the term has been extended to include the physics of the heliosphere (the space around the sun, in principle out to the shock where the solar wind encounters the interstellar medium, but excluding the planets and other condensed bodies).  Here I shall confine my remarks within the original meaning.}\,inferences that have been drawn from, or refined by,  seismology, and the manner in which those inferences have been made, are very briefly described.  Prominence is given to the use of simple formulae, derived either from simple toy models or from asymptotic approximations to more realistic situations, for tailoring procedures to be used for analysing observations in such a way as to answer specific questions about physics.  It is emphasized that precision is not accuracy, and that confusing the two can be quite misleading.
\end{abstract}

\noindent
\section{Prelude}
We have learnt a great deal about the interior of the sun since helioseismology, in the form that we know it, began some 36 years ago. I now take stock of the situation, in an attempt to provide some foundation for asteroseismology, which is already well under way. This is not an attempt to provide a history of the subject, but is instead a few remarks, often in a cautionary vein,  about how one goes about assessing inferences from seismic frequency data.  I shall accommodate what I have to say within a selection of a few investigations that have taught us physics. The details of those investigations do not necessarily apply without modification to other stars, because the data available, both seismic and otherwise, are not of the same kind.  However, in many cases the broad principles behind what I say remain pertinent.

It is not inappropriate to start by describing the first helioseismological inference. It came about from the 
production of a $k$--$\omega$ diagram for high-degree modes by \cite{deubner1975A&A}.   \citet{AndoOsaki1975PASJ}  had already carefully computed a relation from the oscillations of a model of the convection zone, obtaining results very similar to Deubner's observations.  But it was evident that the theoretical frequencies $\omega (k)$ were systematically somewhat too high. In order to revise them downwards it was necessary to produce a model convection zone with a lower adiabatic `constant' $p/\rho^{\gamma_1}$ \citep{DOG1977Nice}, implying a lower entropy for a given chemical composition,  which requires a lesser mean value of the parameter $\Gamma_1 := {\rm {d\, ln} } p /{\rm {d\, ln} }\rho$ defining the seismic stratification through the upper superadiabatic boundary layer. The magnitude of the lessening was estimated, quite crudely, from the analytical dispersion relation for acoustic-gravity waves in a plane-parallel polytrope, and was seen to imply, from an earlier analysis of the influence of the integral properties of that layer on the overall stratification of the convection zone \citep{DOGNOW1976MNRAS}, that the convection zone is about 200\,Mm deep, some 30 per cent deeper than it was fashionable to contemplate at the time.  Soon afterwards this result was confirmed by more precise computations undertaken by \citet{rkuejrdepthconvzone1977ApJ}. 

I tell this story because it illustrates a basic principle that has been used many times subsequently in helioseismology: that to assess the broad implications of a small discrepancy between theory and observation it is adequate -- indeed expedient, in view of its simplicity -- to use at first a very rough description of the possible cause. Of course, more precise -- usually numerical -- analysis is required subsequently in order to quantify the adjustments that must be made to the original reference model, as did Ulrich and Rhodes in the case I am illustrating here. In view of the smallness of the discrepancies between the theoretical eigenfrequencies and those observed, it is usually adequate to presume smallness of the structural adjustment to the theoretical model that is required to remove these discrepancies, and thus for most purposes it is adequate to perform a linearized perturbation analysis. This is the basis of most inversions, which I discuss in the next section. However, some investigators prefer to compare the full frequencies of entire solar models.

I have found the analytical outcome from simple toy models to be extremely useful in exhibiting how properties of the sun depend on details of the representations of the physical processes one is considering.  A prime example is a simplistic approximation to the main-sequence evolution of the solar luminosity, $L(t)$, showing how it depends on mass-loss rate and a putative variation in the constant of gravity $G$ \citep{1990stromgren}.  The formula enables one immediately to determine, for example, the mass-loss rate that renders the luminosity almost constant, a desire amongst climatologists in those days when their theories were incapable of accommodating the inevitable rise in $L$ of mass-preserving stars.  It summarized the published numerical investigations of the day, and, I trust, similar investigations of today \citep[e.g.][]{guzikmussack2010ApJ}, although how well it reproduces the latter has not yet been tested.

An equally valid approach, available only to those with the requisite machinery, is to survey parameter space numerically, recording how various salient properties of solar models respond to changes in the physics.  In a very valuable series of papers \citet{JCDhelioseismology1988, JCDoscillations1991, JCDtestingamodel1996} and \citet{tripahyjcdI_1998A&A...337..579T}
 have published the results of an extensive study in sufficient detail for readers to appreciate how the properties of solar models respond to changes in initial conditions or to assumptions in the  physics on which they depend, and to be able to estimate partial derivatives and thereby carry out multi-parameter investigations for themselves.  In particular, it permitted one to appreciate immediately the implications of the heavy-element-abundance revision proposed by \citet{Asplundetal2005}, a matter to which I shall return later.

A comment related to the implications of the revision in the depth of the convection zone is perhaps not out of place.   Within the framework of standard stellar evolution theory, a deep convection zone at the canonical solar age could be achieved only with a heavy-element abundance, and a consequent initial helium abundance, rather higher than was preferred at the time.  That implied a relatively high neutrino flux, which exacerbated the solar neutrino problem, and heralded the role of seismology in establishing that the solution to the problem must lie in nuclear or particle physics.

In 1970 Fred Hoyle, my Director at the Institute of Theoretical Astronomy, as it was then called, asked me to compute neutrino fluxes from solar models with gravitational settling of heavy elements. I had never before computed a full stellar model, so I resorted to modifying an existing evolution programme which Bohdan Paczynski gave me, adding neutrino production and gravitational settling, the latter rather simplistically by the standards of today. The objective was to determine whether gravitational settling reduces neutrino production. Intuition was not extensive enough to predict the outcome in advance, because almost all intuition of stellar evolution at that time was based on the initial-value problem: how the structure and evolution of a star whose radius and luminosity, say,  are prescribed at $t=0$ responds to a modification to the representation of physical processes. Instead,  the solar problem is a final-value problem, enquiring of 
responses of models whose radii and luminosities are prescribed at  at $t={\rm t}_\odot$; in practice they  are computed by iterating the initial conditions in a series of forward computations.   As is now well appreciated, 
the adjustments at  $t=0$   induce reactions in the opposite sense to those of the modification to the physics, and it therefore takes more careful thinking  to predict the eventual outcome \citep[cf.][]{faulknerswenson1988ApJ}.  Fred was unhappy with my results because the neutrino flux increased, for reasons that we now understand \citep{dog2003Ap&SS}, and the work proceeded no further. But I tell the story because it taught me a lesson in science which I wish to pass on to those not yet experienced enough to have discovered it for themselves. My principal difficulty in carrying out the computations had been to adopt appropriate values for physical quantities arising in the theory, such as cross-sections for the nuclear reactions in the p-p chains: a diversity of values were scattered throughout the literature, and, fortunately, they had not yet been assembled and assessed (and improved).  So I tried a variety, obtaining neutrino fluxes scattered wildly about a value of 20\,snu, a value similar to, although somewhat lower than, the value promulgated by \citet{bahcallsolarneutrinoprediction1964PhRvL, Bahcall15snu1966PhRvL} before Homestake. Then, when almost no neutrinos were detected by \citet{davisharmerhoffmanneutrinos1968PhRvL}, \citet{Bahcall2shaviv1968PhRvL} used `better' nuclear data and produced a best theoretical value of about 7\,snu (still uncomfortably high). When I looked at my results I found that the lowest of my fluxes were near 7\,snu too, and I could go no lower. That was evidently why the neutrino issue became a `problem'. What I learned from the exercise is the manner in which values for uncertain quantities appear to be selected to produce a `best' model. Perhaps appearances are deceptive, but in order to experience them, when it comes to complicated numerical computations, it is necessary to get one's hands dirty by repeating the calculations oneself in order to appreciated the import of published conclusions.  

Even though it has been stressed more than once before, it is still worth stressing again that the oscillation frequencies depend basically on only what I call seismic variables: principally pressure, $p$, exerting a force on material with inertia density, $ \rho$, together with a quantity that relates a (Lagrangian) change $\delta\rho$ to the perturbation $\delta p$ that causes it. The most convenient quantity to adopt for that relation is typically the adiabatic exponent $\gamma_1=(\partial {\rm ln} p/\partial {\rm ln}\rho)_s$, the partial derivative being taken at constant specific entropy $s$.  In a first approximation in which the sun is regarded as being spherically symmetrical, pressure and density are related directly by hydrostatic balance, so only one of them is required to specify the seismic stratification. It is important to appreciate that hydrostatic balance does not depend explicitly on $\gamma_1$, so from a seismological point of view $\gamma_1$ can be regarded as being independent of $p$ and $\rho$,  although of course it must lie within the bounds dictated by thermodynamics.  Any function of $p$, $\rho$ and $\gamma_1$ is also a seismic variable -- most common is the adiabatic sound speed $c=\surd(\gamma_1p/\rho)$ -- and a representation of the seismic variables that is consistent with the seismic data is called a seismic model.  Of course only two independent seismic variables are required to specify the seismic structure completely.

The seismic data that I have in mind in this discussion are the oscillation frequencies of normal modes, for it is they that have been used the most extensively, and they that are the most pertinent to asteroseismology. I should point out that I appreciate that the magnetic field 
$\mathbi{B}$  is also a seismic variable, as also is the angular velocity $\bf {\Omega}$.    Unfortunately, it appears not to be possible to isolate inferences about $\mathbi{B}$  from inferences about $p$ and $\rho$ by seismic frequency analysis alone, because for any stellar model with a given $\mathbi{B}$ there exists an isospectral model with $\mathbi{B} = {\bf 0}$  and an appropriately different $c$; any resulting degeneracy splitting can be represented, at least asymptotically, by a suitable aspherical sound-speed perturbation \citep[e.g.][]{dog1993LH,  egzdog1995ESASP}.  Therefore there is always ambiguity. The eigenfunctions are different, however, although nobody has yet succeeded in detecting and identifying that difference unambiguously in solar observations: to make inferences about $\mathbi{B}$ otherwise requires additional, non-seismic, information.   (Rotation is different, because the distinction between east and west is manifest as a frequency perturbation with a component that is an odd function of azimuthal order $m$, which a magnetic field or a sound-speed perturbation cannot induce.) I hasten to add that the inferences that we have made about the seismic structure, such as those illustrated in Figure 1, do depend also on some non-seismic 
information, namely the values of the seismic radius $R$, which in practice is related to the photospheric  radius via modelling, and the total mass $M$ of the sun.
The seismic radius is the radius at which the variables $p$ and $\rho$, if they were to be extrapolated appropriately from the upper layers of the adiabatically stratified region of the convection zone, would appear to vanish. 
\begin{figure*} \label{fig1}
\centering
     \begin{center}
       \includegraphics[width=13.0cm]{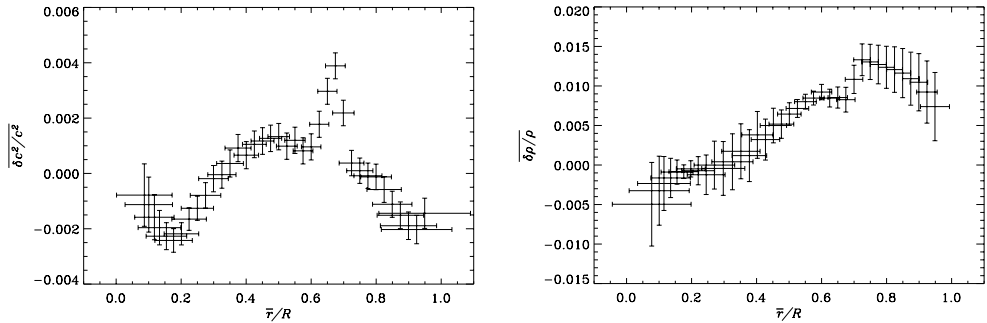}
     \end{center}
\caption{Optimally localized  averages of relative differences between the squared sound speed and  the density in the sun and in Model S of Christensen-Dalsgaard et al. (1996), computed by M. Takata from MDI 360-day data and plotted against the centres $\bar x = \bar r /R$ of the averaging kernels  $A(x;\bar x)$, which here resemble Gaussian functions,  and are defined by $\bar x = \int x A^2 {\rm d}x\,/ \int  A^2 {\rm d}x$.   The length of each horizontal bar is {\it twice} the spread $s$  of the corresponding averaging kernel, defined as $s=12\int(x-\bar x)^2A^2{\rm d}x$ -- an averaging kernel $A$ that is well represented by a Gaussian function of variance $\Delta^2$ has spread approximately $1.7\Delta\simeq0.72FWHM$; were it to be a top-hat function, its spread would be the full width, which is why $s$ is so defined.  The vertical bars extend to $\pm$ one standard deviation of the errors, computed from the frequency errors quoted by the observers assuming them to be statistically independent.}
\end{figure*}
I have implicitly assumed in my discussion that the seismic motion is adiabatic. That is largely true, although not in the immediately subphotospheric regions and the atmosphere immediately above:  the physics in the outer turbulent layers of the sun is ill understood, although fortunately some aspects of the seismic data using modes over a (wider) range of degree $l$ can be used to eliminate the uncertainty.   However, there is much more work that could be done to improve our understanding of the surface, a task which, in my opinion, is urgently required for studying the properties of other stars whose oscillations can be observed at only a few low values of $l$, and for which spatially resolved observations similar to those that have been made on the sun will never be available.  Consequently the effect of the surface layers cannot be unambiguously removed.

\section{On interpreting inversions of oscillation frequencies}
Inversions to determine the seismic structure of the sun are usually carried out by reference to a theoretical model that is sufficiently close to the sun for linearization in the difference to be a good approximation. Here, for simplicity, I assume the sun to be spherically symmetrical, and I represent the structure by two independent seismic variables $y_1(x)$ and $y_2(x)$, where $x= r/R$, $r$ being a radial coordinate. (The procedures can easily be generalized in an obvious way to obtain information about the asymmetric component of the structure). Then one can write, for mode $(n,l)$ of order $n$ and degree $l$:
\begin{equation}
\label{delomegaintegrals}
\delta\omega_{n,l}\simeq\int_0^1 \!K_1^{(n,l)} \delta{\rm ln}y_1\, {\rm d} x + 
\int_0^1 \! K_2^{(n,l)}\delta{\rm ln}y_2\,{\rm d}x + 
{\cal P}(\omega_{n,l})/I^{(n,l)}\,  ,
\end{equation}
where $I^{(n,l)}$ is the inertia of the mode, normalized at $ x=1$, and $\delta$  denotes the difference between the sun and the model. The objective is to express $\delta{\rm ln}y_i$ in terms of the observations $\delta\omega_{n,l}$.  The kernels $K_1$ and $K_2$ depend on $y_1$ and $y_2$ and the eigenfunctions of the mode in question, but not on $\delta {\rm ln} y_1$ and $\delta {\rm ln} y_2$. They can be obtained either by perturbing an integral formula for $\omega_{n,l}$ that constitutes a variational 
principle, if the boundary conditions adopted are such that the system is self-adjoint (the differential operators in the governing equations can be written in self-adjoint form), or, if not, by carrying out a non-singular perturbation expansion of the governing differential system and obtaining $\delta\omega_{n,l}$ from the condition that a solution exists.

The function ${\cal P}(\omega)$ is largely unknown, and was introduced as an acknowledgement that the physics, and therefore the governing differential equations, are not certain in and above the turbulent boundary layer at the top of the convection zone; it might also contain surface integrals which arise when the boundary conditions are such that the system is not self-adjoint. $\cal P$ is a function of $\omega$ alone if the region of uncertainty is thin enough (requiring $l$ to be low enough) for the $l$ dependence of the oscillation to be negligible in that region; otherwise it depends also on $w = \omega/L$, where $L=l + 1/2$, and can usefully be expanded in powers of $w^{-2}$  \citep{dogsvv1995MNRAS}.

The outcome of an inversion is a sequence of estimated spatial averages $\overline{\delta {\rm ln} y_1}$ and $\overline{\delta{\rm ln} y_2}$  of  $\delta {\rm ln} y_1$ and  $\delta {\rm ln} y_2$  weighted by unimodular  (i.e. having unit integral over the domain of existence, here $0<x<1$) averaging kernels $A_1(x;{\overline x})$ and $A_2(x;{\overline x})$ respectively, each of which is a linear combination of $K_1^{(n,l)}(x)$ or $K_2^{(n,l)}(x)$  with coefficients $c_1^{(n,l)}\!$ or $c_2^{(n,l)}\!$ which depend on ${\overline x}$. One would normally like the averaging kernels to be well localized about $x={\overline x}$, for then the averages, which are represented by the same combination of the frequency differences as are the averaging kernels, are relatively easy to interpret:  plotted as a function of ${\overline x}$, they are essentially a blurred view of $\delta {\rm ln} y_1$ and $\delta {\rm ln} y_2$, at least when ${\overline x}$ represents the (actual) centre $\overline x$ of localization of $A_1$ or $A_2$.  (There are some who regard them as tracing the actual functions $\delta {\rm ln} y_1$ and $\delta {\rm ln} y_2$; that would not be a bad approximation if $A_1$ and $A_2$ were very well localized, which can be the case within some ranges of the independent variable $\overline x$.) It should not be  necessary to know how the averaging kernels were constructed in order to interpret the inversions; but it is necessary to appreciate that the final outcome is not uncontaminated: the average of $\delta {\rm ln}y_1$, say, is actually
\begin{equation}
\label{inversionsum}
\overline{\delta {\rm ln} y_1}:= \int{A_1(x;\overline{x})\,\delta{\rm ln}y_1\,{\rm d}x}=\sum_{n,l} c_1^{(n,l)}({\overline x})\,\delta\omega_{n,l}+{\cal R}(\delta {\rm ln}y_2)\, ,
\end{equation}
where $A_1=\sum_{n,l} c_1^{(n,l)}K_1^{(n,l)}$ and the residual is given by 
\begin{equation}
\label{inversionresiduals}
{\cal R} = -\int_0^1 \! \sum_{n,l} c_1^{(n,l)}({\overline x}) K_2^{(n,l)} \delta {\rm ln}y_2\,{\rm d}x - \newline \sum_{n,l} c_1^{(n,l)}({\overline x}) {\cal P}(\omega_{n,l})/I^{(n,l)} \, .
\end{equation}
The corresponding expression for $\overline{\delta {\rm ln} y_2}$ is similar.  It is evident that, 
in addition to obtaining a localized averaging function $A_1$, it is desirable also to reduce the magnitude of $\cal R$ to a (practical) minimum, in order to obtain the best approximation to the average in terms of the the data by ignoring $\cal R$ on the right-hand side of equation (\ref{inversionsum}).  To judge the outcome one needs some information about  the degree to which that reduction has been achieved; such information is rarely available.  One also needs to be informed of the characteristic range of $x$ over which the averaging is taken, and the estimated uncertainty (standard error) of the value of the average that results directly from errors in the data.  That information is usually provided by horizontal and vertical bars, such as those in Figure 1, which respectively represent a characteristic width of the averaging kernels and the standard deviation of the value of the approximated average resulting from the standard errors in the data.  It would be useful also be given an idea of the shapes of the averaging kernels;  some have significant side-lobes far from ${\overline x}$, typically near the sun's surface, demanding some care in interpreting those averages, and possible only if the side-lobe structure is known.  One also needs to know how ${\overline x}$ is defined. If  the averaging kernel is well localized and ${\overline x}$ is some representation of the location of its (suitably defined) centre, then the precise definition is not very important.  But if ${\overline x}$ is merely the value about which the author had tried to locate the averaging kernel, which unfortunately is sometimes the case, then other kinds of information are needed for interpreting the published results.

A common approach to inversion is to try to construct well localized kernels explicitly:  a procedure called OLA (optimally localized averaging).  Greater localization usually results in greater error in the averages arising from errors in the frequency data, because increasing localization increases the magnitudes of the coefficients  $c_i^{(n,l)}$;  more drastic cancellation arises in the sum of the actual frequency differences on the right-hand side of equation (\ref{inversionsum}),  but that is not shared by the random measurement errors.   So a compromise must be made.    Just how that compromise is made depends on the judgement of the inverter.  Once $\overline{\delta{\rm ln}y_i}$ have been estimated,  one can iterate by estimating $\delta{\rm ln}y_2$ from its average $\overline{\delta{\rm ln}y_2}$ (which necessarily requires the adoption of assumptions, such as smoothness, and possibly some prejudices gleaned from one's experience with theoretical models), from which $\cal R$ can then be estimated and incorporated into the constraint  (\ref{inversionsum}). 

Another approach is to try instead to fit the data optimally, using 
parametrized representations of  $\delta {\rm ln} y_1$ and $\delta {\rm ln} y_2$, typically expressed as  linear combinations of preassigned basis functions with coefficients chosen such as to minimize by  weighted least squares the differences $\delta\omega_{n,l}$ given by equation (\ref{inversionsum}).  The outcome is a linear combination of frequencies from which averaging kernels $A_1$ and $A_2$ can be constructed. As with OLA, the procedure must be regularized, usually to favour smoothness, to prevent excessive cancellation and consequent excessive error magnification.  But, unlike OLA, both functions $\delta {\rm ln} y_1$ and $\delta {\rm ln} y_2$ are automatically accounted for simultaneously. This regularized least-squares (data-)fitting (RLSF) method is usually abbreviated as RLS.  When it is used, rather than plotting the averages weighted with $A_1$ and $A_2$, whose side-lobes are  invariably worse  than those explicitly designed by OLA to abhor them, it is common merely to plot the parametrized representations of $\delta {\rm ln} y_1$ and $\delta {\rm ln} y_2$ that result.  These are but a single example of the infinite set of functions that (approximately) satisfy the data.

Inversions are sometimes carried out using asymptotic approximations to the eigenfunctions that yield equations which, after appropriately smoothing the data, can be formally inverted analytically (although the final result must be evaluated numerically) to yield a smoothed representation of the seismic structure. The procedure has the advantage of being simple and fast, and, because the product is an explicit integral, one is readily able to appreciate how certain features in the data relate to features in the seismic structure. Moreover, it does not rely on a reference theoretical model of the sun.  The outcome is a nonlinear combination of the frequencies, so simple averaging kernels are not available.  Inversions such as these are often criticized for being less precise than OLA or RLSF; that they are less precise is indeed often the case, but, by not depending on prejudices such as those upon which any reference model must, they are not necessarily less accurate.  That remark applies particularly to inversions for structure,  which actually depend in a nonlinear way on the functions being sought (and, under some circumstances, in consequence require iteration); it does not apply to inversions for angular velocity $\Omega$, because the sun's rotation is dynamically weak, and the dependence of the eigenfrequencies on ${\Omega}$ can be linearized, leading to  relations of the type (\ref{delomegaintegrals}) (with $y_1={\Omega}$ and $y_2=0$) with kernels $K_1^{(n,l)}$ that do not depend on ${\Omega}$.  It is worth mentioning that it can be convenient to apply asymptotic methods directly to the inversion of the linearized constraints (\ref{delomegaintegrals}) for the structure too, for there it is only the small difference between the sun and the reference model that is being approximated; as I advocated in the first paragraph of the prelude, it is often expedient  to start an investigation with a simple quick analysis, and in some circumstances that analysis can even achieve  adequate precision for the purpose in hand, with the added advantage of a partially analytical appreciation of how the outcome depends on the data.
\begin{figure*} \label{fig2}
\centering
     \begin{center}
       \includegraphics[width=10.0cm]{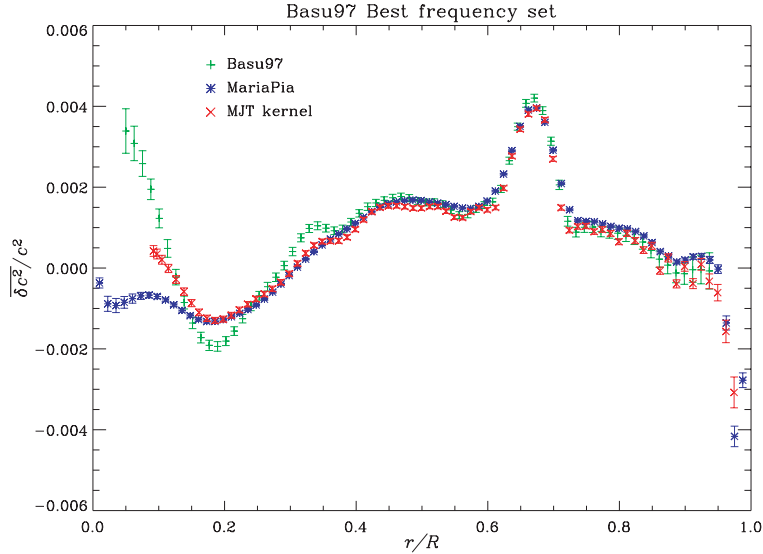}
     \end{center}
\caption{Three different sets of optimally localized  averages of the relative differences between the squared sound speed and the density of the sun and Model S of Christensen-Dalsgaard et al. (1996),  all inferred from the so-called `best frequency set'  selected by 
\citet{Basuetalbestfrequencies1997MNRAS}.   
The vertical bars represent $\pm$ one standard deviation of the propagated frequency errors, assuming those errors to be statistically independent. Characteristic kernel widths have not been drawn for fear of unduly cluttering the diagram (courtesy G. Houdek).}
\end{figure*}
Examples of optimally localized averages of sound-speed differences between the sun and probably the best reference solar model at our disposal, namely Christensen-Dalsgaard's Model S, now an improved version of the model with the same name discussed by 
\citep{jcdetal1996Sci}, are presented in Figure 2.  They were obtained with the same data by several different inverters. 

Despite the abscissa, here labelled $r/R$, not being defined, and likely not being the same for the different plots, the most striking feature is that the differences between the 
differences exceed the quoted errors by a substantial margin:   the accuracy of the results, as naively suggested by the figure, appears to be significantly less than the precision.   It must be 
appreciated, however, that that is actually not the case, assuming that the inverters have not made a technical computational error, which I'm sure they have not.  What must be the case is that the averaging kernels are rather different, and therefore so are the quantities plotted.  One might also note that the tradeoff between kernel spread and error magnification appears to be different from that adopted for Figure1; the errors are smaller here, suggesting broader kernels, which is consistent with the  tachocline anomaly -- the hump in $\overline{\delta c^2}/c^2$ immediately beneath the base of the convection zone which could have resulted from homogenization in the tachocline and which is probably too thin to be resolved \citep{elliottdogsekiitachocline1998ESASP} -- being broader, even  though the data set employed here is different.

An example of the sound speed in another solar model is illustrated in Figure 3. The smallness of the  discrepancy has been employed often by Bahcall,  who, by plotting the error in the theoretical model rather than the solar sound speed relative to a reference model,  presumably, and quite correctly, trusted the seismology more than his modelling.  Indeed, he had   concluded \citep{bahcalltriumphforstellarevolution2001Natur} that the ability to adjust standard solar models  \citep[e.g.][]{brunt-czahnstandardsolarmodels1999ApJ, erratumtobrunt-czahnstandardsolarmodels2000ApJ, BP2000standardsolarmodels} to bring their neutrino fluxes within a mere 20\%  or so of those measured at SuperKamiokande and the Sudbury Neutrino Observatory (\citet{ahmadetalneutrinos2001PhRvL..87g1301A}, cf. \citet{ahmadetal_a_2002PhRvL..89a1301A}) demonstrates `a triumph for the theory of stellar evolution ... which is a cause for rejoicing among astronomers'.
\begin{figure*} \label{fig3}
\centering
     \begin{center}
       \includegraphics[width=8.0cm]{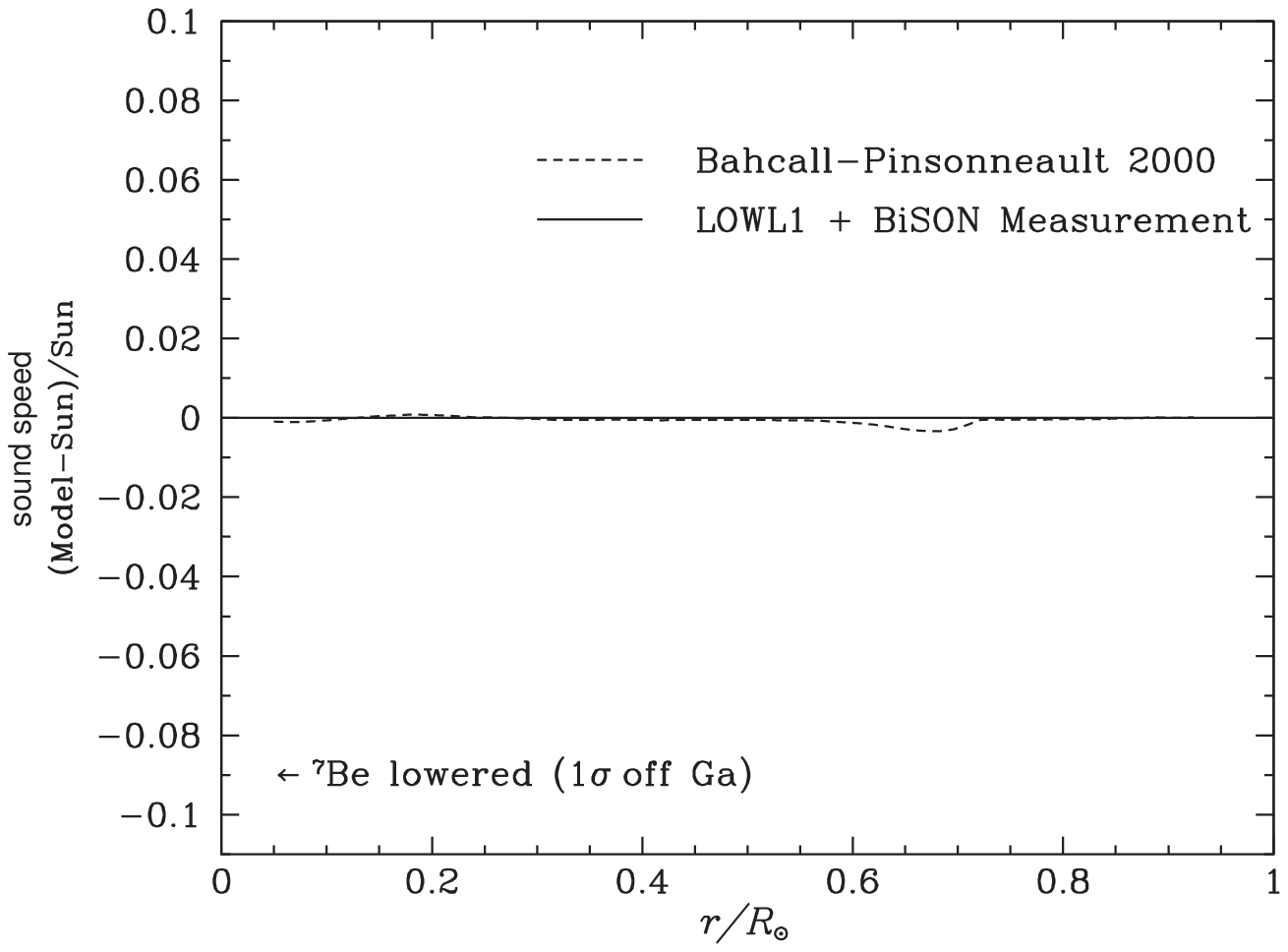}
     \end{center}
\caption{The dashed line is the relative difference between the sound speed in a standard solar model computed by Bahcall and Pinsonneault and that inferred by Basu \citep{Basuetalbestfrequencies1997MNRAS} from a combination of oscillation frequency data obtained by \citet{chaplinetalfrequencies1996SoPh.168.1C} and \citet{tomczykschoumjt1995ApJ...448L..57T,tomczyketalLOWL1995SoPh..159....1T}  \citep[from][]{Bahcallpinsonneaultbasu2001ApJ}.}
\end{figure*}

\section{Macroscopic Physics}
The most fertile arena of macroscopic physics has arisen from studies
of kinematics\footnote{Invariably, in the present context, called dynamics,
although no direct helioseismological measurement of a force has ever
been made.  However, the kinematical inferences have spawned a great
deal of theoretical research into the dynamics.}.  The most 
plentiful, and the most reliable, are measurements of rotation.  The reason is
partly that the signature in the oscillation frequencies is greater
than those from other components of flow, but principally because
it is only rotation that is uniquely identifiable:  by the property
that eastward and westward (azimuthally averaged) macroscopic motion would 
split the degeneracy of the seismic frequencies with respect to
azimuthal order $m$ in opposite senses.  The splitting is caused
by both advection and a Coriolis force, and is a function that has an
expansion in powers of $m$ with  non-zero coefficients of both odd and even powers.  
An expansion
of frequency splitting produced by any other (aspherical) perturbing agent
contains only even powers of $m$.

Plots of the time-averaged angular-velocity averages 
$\overline{\Omega}(r,\theta )$ are presented in other contributions to
these proceedings, so I refrain from doing so too. The dominant
features are the (principally latitudinal) differential rotation in
the convection zone, the almost (perhaps exactly) uniform rotation in
almost all of the radiative interior, together with the thin tachocline
separating the two.  The tachocline is now amongst the most active
arenas of heliophysical research \citep[e.g.][]{hughesrosnerweisstach2007}.  The rate of rotation of the radiative zone is such that the
spherically averaged linear velocity suffers a 7\% decline at the equator going
downwards across the tachocline.  

That the angular velocity deep in the
interior is not much greater than that observed at the surface was
realized in the early days of helioseismology \citep{Naturerotation1984Natur}.  That was a surprise to
many theorists. It had generally been believed that, as a result of
the retarding torque applied via Maxwell stresses to the sun by the
solar wind, the surface layers must now be rotating substantially more slowly 
than the deep interior;  debate was about only by how much.    \citet{dicke1964Natur},
for example, maintained steadfastly that the difference is substantial, justifying his
position with his surface oblateness measurements with Goldenberg  \citep[see][]{DickeGold1967PhRvL, dickegoldenberg1974ApJS}, and arguing that viscous stresses in the radiative interior
were insufficient to tie the core to the surface. He used the assertion to support his theory of gravity
with Brans \citep{BransDicke1961}: very roughly speaking
the theory regarded Newton's constant $G$ not as a true constant,
but as a field which satisfied an inhomogeneous wave equation fed by matter -- gravity was thereby
less tightly connected to matter, and therefore the precession of
planetary orbits, for example, must be slower than what had been
predicted by General Relativity;  appropriate rapid rotation of the
solar interior was therefore required to distort the external
gravitational field enough to make up for the planetary precessional
loss.  The helioseismological finding destroyed that argument.

Dicke's weak-spin-down claim triggered more cogent fluid-dynamical
discussion, principally by \citet{howardmoorespiegel67}   and \citet{brethertonspiegel1968ApJ}, who pointed out that the baroclinicity induced by the differential
rotation of spin-down drives a
global meridional flow which transports (negative) angular momentum
downwards from the convection zone, thereby slowing the core much more
quickly.  The process was likened to that operating in a stirred cup
of tea, in which the tea slows on a timescale equal to the geometrical
mean of Dicke's global (viscous) diffusion time and the period of
rotation\footnote{The analogy is not perfect,  because, unlike a cup of tea, the sun is thermally stratified, and spin-down is moderated via thermal diffusion, adding richness (i.e. complicating) the analysis \citep[e.g.][]{dickespindown1967ApJ}.  But
the general principle remains;  the basic dynamical processes are not completely prevented from operating.  A recent quantitative investigation of the processes involved  has been presented 
by Spiegel and Zahn (1992) in another context.}.

Interestingly, it was \citet{einsteinmeanders} who first used such an
argument, in explaining the meanders of rivers, not realizing at the time that he was establishing ammunition to be used  much later in defence of his General Theory of Relativity.

The detailed knowledge we now have of $\overline{\Omega}(r,\theta,t)$,  
coupled with our knowledge of the seismic hydrostatic
stratification, enables us to calculate the shapes of the
gravitational equipotentials more accurately than by any other means available 
today.  They can be expressed in terms of the (even) coefficients
$J_{2 k}$ of a multipole expansion. The quadrupole moment, 
$J_2 = 2.2 \times 10^{-7}$ \citep{schouetalrotation1998ApJ, antiachitregoughsolarKE2008A&A}, contributes the most to the precession of
the perihelion of the orbit of Mercury.  That value induces an orbital precession which, with the current  measurement precision, is too small to influence the precessional test of the theory of General Relativity.

The baroclinicity induced by the (latitudinal) differential rotation
of the convection zone also drives meridional flow, which, unless
strongly opposed by some agent, would transmit the latitudinal
variation of $\Omega$ into the bulk of the radiative interior on a
timescale less than the age of the sun \citep{easjpztach1992A&A}.
Notwithstanding Spiegel and Zahn's hypothesis that appropriate
anisotropy of turbulence induced by instability of that flow would
confine the rotational shear to a tachocline (a process challenged by
\citet{elliott1997A&A}), McIntyre and I (1998) have argued that no
purely fluid-dynamical process
can maintain the uniformity of $\Omega$: in the radiative interior the
only remaining possibility is a presumably fossil, and therefore
no doubt predominantly dipolar, magnetic field.  Because magnetic
diffusivity is so small, the  (horizontal component of the) field
is prevented from penetrating the tachocline by the downwelling
tachocline flow, which occurs at all latitudes except those at which
rotational shear is negligible (near latitudes $\pm$ 30$^\circ$).  I
still believe this to be the case, despite the counterclaim by \citet{brunzahn2006A&A}, 
which they tried to support with (necessarily
excessively diffusive) numerical simulation.  Garaud and her colleagues \citep[e.g.][]{garaudtachoclineI2002MNRAS, garaudgaraud2008MNRAS, garaudacevedoarreguin2009ApJ} 
have carried out a series of calculations with lower diffusion coefficients, but at the
price of assuming axisymmetry; and \citet{woodmccaslingaraud2011ApJ} 
have gone a long way towards demonstrating the case.  But there have been 
problems with preventing field penetration of the tachocline near the
poles, where the dipole field is vertical and where the tachocline
circulation has almost no horizontal component to sweep it aside, notwithstanding the demonstration by \citet{tswmemtach2011JFM} of the existence of a steady state with the field confined to the radiative interior, even when it is symmetric about the rotation axis.  I believe
that that particular problem is due at least partly to the superficially simplifying
assumption of axisymmetry.  As has been discussed elsewhere \citep{DOG2012GApFD},   
were the axis of the magnetic dipole
initially not to have been aligned with that of the angular velocity
in the tachocline, the tachocline circulation is likely to have applied a torque between the 
dipole and the convection zone in such a sense as to cause the dipole axis to migrate towards the latitudes of zero tachocline shear.  Unfortunately the magnetic field appears to be  too
weak for that to be detected directly by seismology.

In addition to the magnetic field impinging on the tachocline from
beneath, there is also the possibility of the field being pumped or
diffused into the tachocline from above.  That field is likely to
change sign with the solar cycle, and hence decay in a distance much
less than the tachocline thickness \citep{pascalebpropagationt1999MNRAS}, even in the
presence of the baroclinically driven tachocline downwelling flow of
the magnitude inferred by \citet{dogmem1998Nature}.

An important consequence of the tachocline circulation is that it mixes back into
the convection zone
helium and heavier elements which tend to settle under gravity.  That process reduces the mean molecular mass $\mu$ of the
material in the tachocline, and thereby increases the sound speed.
That, I am sure, is the explanation of the tachocline sound-speed anomaly
evident in Figures 1 and 2 as a hump between $x = 0.6$ and 0.7 in the
sound-speed excess over that in Christensen-Dalsgaard's theoretical
Model S.  Note that it actually represents a smoothing of the sound
speed.  I maintain also that it is probably the meridional circulation
that is the primary smoothing agent, notwithstanding the possibility
of additional small-scale shear turbulence or convective overshoot.
Think of the Gulf Stream, which transports, primarily by advection,
heat from the Carribean to the coasts of north-western Europe.  The
associated turbulent transport is too weak to compete.  Then notice
that the Richardson number $N^2/(\Delta\Omega)^2$ associated with the rotational shear  in the
tachocline (which is $10^{12}$ times the Richardson number associated with
the tachocline circulation) -- about $3 \times 10^{6}$ -- is some hundred 
times more than that pertaining to the Gulf Stream.  So the sun
appears to be much more stable, and is unlikely to be subject to more
intense hydrodynamically driven turbulence than is the Gulf Stream. However, there does
remain the possibility of magnetorotational instability, which is
difficult to assess because the configuration of any weak magnetic
field that might be present in the body of the tachocline (which is
essentially nonexistent in the picture I have just painted) is not
known.

The physics of the remaining, large-scale, difference between the
sound speed in the radiative zones of the sun and Model S has not been
convincingly identified.  There are a variety of possibilities, some
of which I shall address below.  The form of the sound-speed in the
adiabatically stratified region of the convection zone of the sun is
reasonably well established:
$c^2 \simeq \left(\gamma_1 - 1 \right) GM \left(r^{-1} -R^{-1}\right)$, 
where $M$ is the mass of the sun out to the radius $r$ (assumed constant for deriving this approximate relation) and $\gamma_1 \simeq $ constant, so it appears that
the discrepancy must result principally from adopting the wrong value
of the seismic radius $R$.  However, that does not explain the entire
discrepancy;  the complete resolution may be related to asphericity at
the base of the convection zone, although I hasten to add that the
base of the tachocline is almost certainly spherical, because the
magnetic field is not strong enough to support any significant
asphericity against the $\mu$ gradient in the radiative interior.

Returning to the angular velocity, I believe that other features of
the helioseismological inferences, such as the subphotospheric shear
\citep{schouetalrotation1998ApJ}, the torsional oscillations \citep{vorontsovetaltorsionalosc2002Sci} and the tachocline oscillation  \citep{howeetaltachoclineoscillations2000Sci, howeetaltorsoscupdate2011JPhCS}, which, with
eye of faith, might also be discernible another half-wavelength or so
deeper into the radiative zone, at least before the oscillation disappeared in 2000, do not have simple cogent  explanations -- with clarity sufficient for me to explain them to my grandchildren 
-- although they are rightly  (except, perhaps, the so-called
tachocline oscillation) subjects of current research.  However, the
seismic inferences concerning $\Omega$ are providing an invaluable
reference to guide the very fruitful numerical simulations of solar
convection discussed by Toomre in these proceedings.

\section{Microscopic Physics}
The microscopic physics that might be accessible to seismological investigation concerns  principally the adiabatic exponent $\gamma_1$ and the thermally pertinent quantities $\kappa$ and $\varepsilon$: opacity and the rate of generation of heat by nuclear reactions. The second and third are evidently accessible only with the help of non-seismic information, because they involve temperature, which is a non-seismic quantity. Moreover, investigation of the first also requires non-seismic augmentation, because there is no redundancy in the equations governing adiabatic seismic oscillations.  In all cases it is necessary to consider properties of theoretical stellar models, deriving from them constraints (always subject to the adoption of certain assumptions) that can be imposed upon seismological inferences.

Adopting an equation of state that delivers $\gamma_1(p,\rho;X_i)$, where $X_i$ are the abundances of the chemical elements, enables one to estimate, at least in principle, those abundances seismologically, using the depression of $\gamma_1$ by ionization.  Provided that the domain of investigation is deep enough in the convection zone where we believe the stratification to be adiabatic -- itself a non-seismic constraint -- the consequent relation between the variation of $p$ and $\rho$ restricts ambiguity in the magnitude and form 
of the depression, thereby permitting a calibration of $X_i$. To date, only the abundance of helium has been reasonably reliably estimated (the hydrogen ionization zone lies  in the superadiabatic boundary layer whose structure is uncertain because it depends on the treatment of convection, and therefore the depression of $\gamma_1$ cannot be measured) -- the $\gamma_1$ depression resulting from the ionization of individual heavier elements is too small to measure -- although measurement of a combined depression is planned for estimating the total heavy-element abundance by accepting relative abundances obtained from spectroscopic studies of the solar atmosphere \citep{kmdog2009}. This could be fraught with uncertainty, because recent modifications to the relative abundances \citep{GrevesseAsplundSauval2011sswh.book, Caffauetal2011SolPh}, whose effect on opacity I mention at the end of this section, are still in some doubt.   Additionally, an estimate based on the spatial mean value of a diagnostic thermodynamic function $\Theta(\gamma_1;r)$ which responds to ionization  has been undertaken \citep{antiabasuZdetermination2006ApJ}, but that procedure relies on the absolute value of $\Theta$  rather than its local deviations, and is therefore much more susceptible to uncertainties in  the equation of state \citep{BaturinWDDOGSVV2000MNRAS}:  the reliability of the value of $\gamma_1$ obtained from modern equations of state has been discussed extensively \citep{JCDWDEoS1992A&ARv, dappenlebretonrogersEoS1990SoPh, dappenEoS1998SSRv, dappenanyfonovEoS2000ApJS}, and was questioned by \citet{basujcdintrinsiceos1997A&A} and by \citet{basudappennayfonov1999ApJ}, who attempted a seismological inversion for $\gamma_1$,  suggesting that the uncertainties are as great as the heavy-element induced depression itself.

For the sake of the unwary reader, I draw attention to the fact that the questioning has not   completely been answered, because is was not possible to eliminate unwanted integrals such as that in $\cal R$ on the right-hand side of Equation (2) to isolate $\delta {\rm ln}\gamma_1$, an inevitable consequence of the lack of redundancy in the oscillation equations which I mentioned earlier, and which render it logically impossible to determine the intrinsic error in $\gamma_1$ by seismological means alone. Some progress was made later by \citet{rabellosoaresetalintrinsicgamma_12000SoPh} and \citet{dimaurobasuetalintrinsicgamma_12002A&A}, in which the functional form of the seismologically inaccessible error of $\gamma_1$ was estimated by implicitly using what I presume was assumed 
to be a more robust 
aspect of that same equation of state.   It was concluded that quite good estimates of the uncertainty in $\gamma_1$ could be obtained in regions in which $(\partial {\rm ln} \gamma_1/ \partial  {\rm ln} Y)_u$ is small (here $u=p/\rho$ is the square of the isothermal sound speed), but not in the helium ionization zone where it is not (and where a reliable equation of state is needed for a sound determination of the helium abundance). Therefore there remains some uncertainty in direct seismological estimates of the helium abundance.

 I judge from the modern literature that the initial (ZAMS) helium abundance of the sun has been determined to be $Y_0=0.25\pm0.01$, the `errors' being estimates of accuracy, not precision. Interestingly, this is the same as the value estimated from early seismic model-fitting \citep[e.g.][]{DOG1983PhysBull}, although in those early days the precision was only half as good as it is today, and the uncertainty was probably rather greater.  
 
  I have spoken  of the adiabatically stratified region of the
convection zone as though its existence is obvious.  But should that not
be checked?  To be sure, the interiors of the high-Rayleigh-number
convection zones with which we are more familiar, such as occur in some laboratory experiments and in
clouds in the Earth's atmosphere, appear to be adiabatically
stratified, and modern numerical simulations such as those discussed by
Toomre in these proceedings exhibit that property too.  But should such arguments be trusted? 
As a pertinent aside I might remark that the original motivation for Davis's neutrino observatory  was to confirm what most astrophysicists took for granted:  that the sun really is
powered by nuclear transmutation, even though the 
argument by \citet{eddingtoninternalconstitution1926ics} seemed invincible, if not when it was first
propounded.  Much effort was subsequently expended in checking the details.  
Likewise, it should perhaps be considered a worthy endeavour to investigate 
convective stratification more thoroughly.  The difficulty in checking it 
seismologically is that its effect on the propagation of acoustic waves is only 
via the influence of buoyancy, which is tiny and therefore necessarily
limits precision severely.  So far as I am aware, there has been only one
attempt to test the stratification, yielding $|\gamma_1^{-1} - \Gamma_1^{-1} | \simeq |\nabla - \nabla_{\rm ad} | \simlt 0.03$ \citep{DOGCatania1984MmSAI},
 which, given that mixing-length theory
predicts values of order $10^{-6}$ in the lower half of the
convection zone, may not seem a very tight constraint.
It would be interesting to repeat the exercise with modern
data using a direct inversion similar to that described by \citet{elliottseismiceos1996MNRAS}.

I come now to opacity. Early sound-speed inversions \citep{speedofsoundJCDetal1985Natur} suggested that opacity computations of the day were about 20 per cent too low immediately beneath the convection zone down to temperatures of about $4\times10^6$\,K. Subsequent scrutiny \citep{iglesiasrogerscepheidopacity1990ApJ, Iglesiasrogersopacity1991ApJ} revealed an error in the treatment of spin-orbit coupling in the radiative-transition calculations that had been carried out at the Los Alamos National Laboratory, and some other, more technical, matters into which I shall not delve here. The error was found to affect $\kappa$ by even more at lower temperatures, not relevant to the sun because they occur in the convection zone.   Correcting $\kappa$ resolved several important issues in astrophysics, such as  the excitation of $\beta$ Cephei and SPB stars \citep{moskalikdziembowskibetacephei1992A&A, DziembowskietalSPBstars1994IAUS}.  With the new opacities, and other improvements such as the incorporation of gravitational settling against diffusion, the superb Model S of \citet{jcdetal1996Sci}   was constructed; it has remained 
the most well-used reference model ever since.

Opacity became a hot topic again following the new
spectroscopic abundance analyses by 
\citet{Asplundetal2005, Asplundetal2009ARA&A}, \citet{GrevesseAsplundSauval2011sswh.book} and  \citet{caffauetal2009A&A, Caffauetal2011SolPh},
in which three-dimensional hydrodynamical solar atmospheric models of 
\citet{steinnordlundatmosprops1998ApJ} \citep[see also][]{nordlundsteinasplund2009LRSP}  
and \citet{caffauetal3dmodel2008A&A} 
were used instead of one of the usual one-dimensional essentially hydrostatic
models.  The surprise was that the abundances of the opacity-producing
elements C, N and O were first reported by Asplund and his colleagues to be about 30\% lower than
previously believed, although the values have risen somewhat since;  those reported by Caffau and his colleagues are somewhat higher still, although low compared with the older values of \citet{grevessesauvalabundances1998SSRv}.    Although the photosheric abundance of Ne, the remaining substantial
contributor to opacity, cannot be measured accurately,  it seemed 
likely that the effect of the convective fluctuations on the spectrum
lines influence the abundance analysis similarly (thereby laying doubt on the suggestion by \citet{draketesta2005Natur} that the abundance of Ne is very much greater than the value normally adopted -- see also \citet{YoungFIPquietsun2005A&A, Asplundetal2009ARA&A}).   This posed a problem.  The effect of varying opacity in solar
models was already known to change the sound speed nontrivially;
indeed, in an early series of papers Christensen-Dalsgaard
demonstrated that $\delta c^2/\delta\xi$, where $\xi$ is almost anything, is
nonzero, and quite different for different parameters, or functions,
$\xi$ -- sufficiently different that it is unlikely that they are not
linearly independent, and so adjusting other properties of the models could not plausibly be contrived to cancel the opacity discrepancy. Therefore Asplund's result destroyed the
apparently superb correspondence of the seismic variables of Model S with
reality.  It was immediately obvious that something must be done to
restore the opacity.  

The problem posed by Asplund within the framework of standard
solar-evolution theory is easily understood.  Consider first the known
seismic structure.  Then note that the variation of the surface
luminosity $L(t)$ is insensitive to assumptions about the internal
structure.  Consequently $\int\! L\, {\rm d}t$ is well determined -- given that
we (think that we) are pretty sure of the sun's age --  and so therefore is the total
amount of hydrogen that has been consumed\footnote{To be sure, there are
variations amongst models in the balance of the ppI and ppII chains
which modify the total somewhat, but that is small compared with the
enormous change in the value of the total heavy-element abundance $Z$
that we are addressing: so too is the effect on the equation of state.}. 
We can therefore safely take 
the absolute deficiency in the hydrogen abundance $\Delta X(r,t)$ to
be the same as that in Model S, and hence obtain from $p$ and $\rho$ the
temperature $T$ in terms of the (presently unknown) initial hydrogen
abundance $X_0$.  From the outcome can be calculated the total rate of
generation of energy by nuclear reactions, from which $X_0$ can be
determined by equating that rate with the observed luminosity.  To be
sure, the last step depends on accepting nuclear-reaction cross
sections, but after decades of investigation by those in pursuit of a
resolution to the solar neutrino problem I recommend that they be
accepted, at least for the time being.  One now has all the quantities present in the radiative
transport equation, save the opacity $\kappa$.  Hence $\kappa$ can be evaluated.  
The difference between that and the opacity in Model S is
plotted in Figure 4.  The problem posed by Asplund is simply to
reconcile that function with his surface abundance measurements, which seem to imply opacities that differ from the those in Model S by some 30 per cent or so.
\begin{figure*} \label{fig4}
\centering
     \begin{center}
       \includegraphics[width=9.0cm]{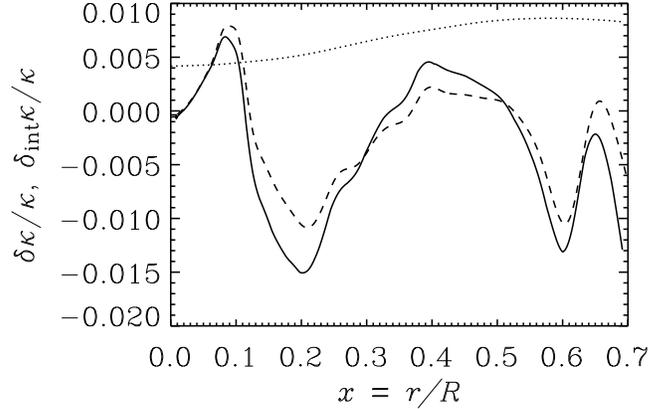}
     \end{center}
\caption{The continuous line is the relative difference between  averages of the opacity $\kappa$ in the sun and the opacity in Model S of Christensen-Dalsgaard et al. (1996), inferred from optimally localized averages of sound-speed and density differences coupled with an estimate of the helium abundance $Y$ and the assumption of thermal balance throughout, as described in the text.  The dashed curve is the corresponding difference along the inferred $\bar \rho - \bar T - Y$ path in the sun, plotted against  the centres of the averaging kernels.  The dotted curve is $0.01 (\partial {\rm ln} \kappa / \partial {\rm ln} Z)_{\rho,T,X} $  \citep[from][]{DOGLorentz2004AIPC}.  }
\end{figure*}
Of course one could instead have taken a different route by accepting
the opacity and computing the nuclear energy generation rate
$\varepsilon$, but the outcome of that would obviously have been to find
regions in the sun in which $\varepsilon < 0$.  Surely that also supports
my recommendation not to do so.

In attempts to shed light on the matter, many papers were written in
which properties of solar models were changed, some accompanied by
unnecessary inversions,  reiterating what Christensen-Dalsgaard had taught us in the past;  
\citet{basuantiaabundancereview2008PhR} 
have compiled a useful catalogue.   However, some were novel.  If one sets aside the idea that the new abundance estimates are too low, despite the care that has gone into deriving them, the paper that stands out 
from the rest in my mind is by \citet{guziketalsolarcontamination2005ApJ}, who cut the cackle\footnote{\citet{J2, Trewin67}} by rejecting standard
solar-evolution theory and posited that the chemical composition in
the sun's radiative zone is essentially no different from that in
Model S.   How could that be? The hypothesis is that, after most of the sun had condensed
from the interstellar medium, it accreted metal-deficient material,
obscuring its true colours.  Such accretion could hardly have occurred
well into the main sequence, since the accreted material had to be gaseous,
and would have been inhibited by the solar wind.  But, granted that
the proto-solar accretion disc  was inhomogeneous, with incipient
planetary condensations into which solid grains were preferentially
drawn, couldn't the largest of them have actually seeded the 
sun, onto which some normal, somewhat metal-deficient, gaseous disc material subsequently
accreted?  A somewhat modified story was entertained recently by \citet{solarlowZcontamination2009ApJ}: accretion onto the early sun of  dust-cleansed proto-planetary nebular material, essentially what \citet{guziketalsolarcontamination2005ApJ} had had in mind.  In either case, the chemical inhomogeneity that remained in the sun would
be Rayleigh-Taylor stable, and stable also to fingering, and would 
plausibly have created a discontinuity in composition, beneath the present-day base of the convection zone,  that has
survived until today.  If so it would have seismological consequences: it
would produce an oscillatory signature in the eigenfrequencies not
unlike that produced by the abrupt changes in stratification at the
base of the convection zone that we model already.  Its amplitude would be no greater than
about 25\% of that of the convection-zone signature, and the two would
be entangled, rendering unambiguous detection difficult.  Nevertheless, it is
worth looking for.

In a similar vein to my preferred statement of the abundance problem, \citet{JCDetalopacity2009} have recently asked the question: by how
much would the opacity formula need to be changed were the chemical
composition in the radiative interior to be (after accounting for
gravitation settling, of course) changed to be consistent with
the determination by \citet{Asplundetal2009ARA&A}?  The
outcome is an almost linear function of $\log T$ (along the thermodynamic
$\rho-T$ path of Model S), declining from about 30\% at the base of the convection zone to 
about 6\% at the centre.  One might have thought that it should be simply 
$-\sum_i (\partial \kappa / \partial X_i)\delta X_i$, once again,
computed along the $\rho-T$ path of Model S, where $\delta X_i$ are the
Asplund-Grevesse modifications.  I tried approximating the opacity
modification by $(\partial \kappa / \partial Z)\delta Z$ assuming the
relative abundances not to have changed, and was somewhat surprised to obtain a
rather different result: my modifications were less close to being a linear function of ${\rm log}T$, being some 5\% greater than the values obtained by \citet{JCDetalopacity2009}  immediately beneath the convection zone, and about 2\% less  interior to $r/R=0.5$.   That caused me to wonder whether I had made a mistake in my simple calculation.  However, during my presentation at the Fujihara seminar Christensen-Dalsgaard assured us that his result deviates from my simple estimate because in his calculations the relative abundances of the opacity-producing elements had also been modified.  That hadn't been clear to me when I read the paper.    Is this an example of where simple calculations go awry? Yes, if one takes the results too seriously without assessing the precision of what is being done, or without being absolutely sure of what one is comparing the outcome with.  So here is another lesson to be learned. 

\section{Seismic model-calibration}
Calibrating theoretical solar models against seismic data was the
first means by which inferences about the inner state of the sun were
drawn.  I have already mentioned in the prelude to this contribution
the calibration of the upper superadiabatic convective boundary layer,
first from an analytically estimated perturbation, then by direct
comparison of the full oscillation frequencies  of a set of
numerically computed envelope models.  That led to the first seismological revision 
of the depth of the convection zone, and then, using that result as a
further calibrating datum for full solar models, to a seismic estimate
of the initial helium abundance $Y_0$ \citep[e.g.][]{DOGprotosolarY1983}.
The location of the base of the convection zone has been used
extensively as a datum for assessing or calibrating solar models since \citep[e.g.][]{bahcallbasupinsonneault1998PhLB}, although there are exceptions.  Nowadays, more
extensive, and often more highly processed, seismic data are 
used,  including other aspects of entire seismic models, to assess or calibrate 
evolved solar models \citep[e.g.][]{Turckchiezeetalseismicmodel2001ApJ}\footnote{Turck-Chi\`eze {\it et al.} called the seismically
calibrated model a seismic model, notwithstanding its strong dependence on
nonseismic argument.  That is contrary to the usage of the term in
this discussion, and to common usage in both helioseismology
and geoseismology.   I remark that RLS data-fitting could also be regarded
as a (functional) calibration procedure, but because no non-seismic
constraint is overtly imposed (other than the values of $M$ and $R$ -- the
regularizing penalty function is apparently arbitrary, although its
definition is seismically motivated), it is reasonable in that case to call the
outcome a seismic model.}.

The first overt global seismic calibration of entire solar models was simply a
naive least-squares frequency fitting \citep{JCDDOG1981A&A}.  Interest was principally in the abundances of helium
and heavy elements, and the consequent implication concerning neutrino
production, so only low-degree modes which penetrate into the radiative
zone were used.  Several local minima in $\chi^2$ were found
(depending on the values adopted for the orders $n$ of the modes that were used, for they were not known at the time), the
best two having helium abundances $Y$ of a little above and substantially
below 0.25 (a commonly favoured value of the day), and corresponding
heavy-element abundances above and below 0.02.  The helium-rich fit was
somewhat better, although perhaps not significantly so.  However, if
one coupled those frequency fittings with the earlier seismological finding from
high-degree modes that the convective zone is deeper than previously
preferred, then the helium-rich solution was definitely favoured.
That appeared to establish that the neutrino problem was an issue for nuclear or
particle physics, not directly a problem for global heliophysics.  Subsequent full inversions 
of the kind illustrated in Figures 1 and 2 were required to put the final nail in the coffin.  
What was significant was that the
frequency residuals of the best-fitting model were substantially greater than the 
standard errors in the data.
Thus it was evident that none of the models adequately represents the
sun.  That does not imply that the models are of no use to address
specific scientific questions:  rather than mindlessly trying to fit
all the data together, one should try instead to extract from them signatures
that are sensitive to the matter in question, and insensitive to
extraneous properties.

Calibrations come into their own when the matter in hand is beneath
the resolution of straightforward inversion.  An important example is
the thickness of the tachocline. 
As originally conceived by \citet{eastachycline1972NASSP} and \citet{easjpztach1992A&A}, the tachocline is
the transition region beneath the convection zone and the uniformly
rotating radiative interior.  It is the region in which material is
homogenized with the convection zone, producing the sound-speed
anomaly evident in Figures 1 and 2 (and the corresponding
near-discontinuity in density).  Therefore, its thickness can be
determined by calibrating the magnitude of the sound-speed anomaly 
\citep{elliottdogsekiitachocline1998ESASP}.  
The procedure is precise, but the
accuracy of the outcome depends crucially on the accuracy with which
gravitational settling has been taken into account in the reference
model, and, probably to a lesser extent, on the structure of the
mixed layer.  Alternatively, one can take the name literally, as have
\citet{AGKtachocline1996ApJ}, \citet{charbonneautachocline1998JRASC} and \citet{charbonneauetaltachocline1999ApJ}, and try to measure the
extent of the rotational shear by calibrating a plausible
parametrized function (there is yet no reliable theoretical prediction
of the functional form) against an inversion for angular velocity. 
 
A related calibration concerns the form of the shear itself.
Straightforward inversions provide only a smooth variation with
latitude.  Yet if the magnetic field is dragged into the convection zone
at mid latitudes by the upwelling tachocline circulation that \citet{easjpztach1992A&A} and McIntyre
and I (1998) have described, shouldn't the consequent Maxwell stresses
create a finite region of zero shear?  Sekii and I have designed a
putative seismic signature for detecting such a region, and we
hope to put it to use when next we have adequate time together. 

I conclude my discussion of this topic by addressing what I regard as
a major calibration.  It is designed to determine how much hydrogen
has been consumed by nuclear reactions throughout the lifetime of the
sun in order to estimate the sun's age.  The long-term goal of this
continuing investigation, which we admit might be pie in the sky, is
to ascertain whether, and if so by how long, the meteorites condensed
after the formation of the sun.  The principle of the calibration is
to use a signature of low-degree p-mode frequencies that is sensitive
particularly to the stratification of the core.  The stratification
evolves with time, in a manner that depends on the proportion of
hydrogen -- and therefore helium -- in the core \citep{JCDdiagram1988IAUS, DOGlinearXc-t1995ASPC}.   Therefore, it is necessary to ascertain
the absolute helium abundance too.  An early discussion by  \citet{dziembowskietalsolarage1999A&A} 
ignored the abundance issue.  I tried a two-parameter ($t_\odot, Y_0$)
calibration \citep{solaragevhippel2001ASPC} using for data two values of the so-called
small frequency separation $d_{n,l} := \nu_{n,l} - \nu_{n-1,l+2}$ averaged over different
domains of ($n,l$).  The small separation is most sensitive to changes
in the core; although, formally, $d_{n,l}$ depends just as much on the
stratification of the rest of the star \citep[e.g.][]{DOGnovotny1990SoPh},
the stratification outside the core  hardly varies with $Y_0$ -- or, equivalently, $Z_0$ -- and
$t_\odot$, and it was hoped that the two different averages would
weight $t_\odot$ and $Y_0$ sufficiently differently to enable them to be distinguished.  That turned out to be only marginally possible, 
as becomes evident by comparing the integrands for $d_{n,l}$ as $t_\odot$ and $Z_0$ are varied (illustrated by \citet{DOGnovotny1990SoPh} and \citet{GHDOGsolarage2011MNRAS}, respectively).  Therefore the calibration, which yielded $t_\odot=4.57$Ga, and a rather high $Z_0$,  is uncomfortably susceptible to
frequency and modelling errors.  Soon afterwards,  \citet{bonannoschattlpaternosolarage2002A&A} repeated the more robust single-parameter
calibration of Dziembowski et al. for $t_\odot$ by simply adopting a plausible value of $Z_0$,  obtaining the same age as  \citet{solaragevhippel2001ASPC} .    Subsequently 
Houdek and I (2011) developed a more robust
two-parameter procedure, using the oscillatory signature of helium
ionization \citep{DOGOji1990LNP, HGseconddiffs2007MNRAS} in low-degree p modes to measure $Y$.  Our
result, which we hoped to be more reliable than what is obtained from a single-parameter fit,  is $t_\odot = 4.60 \pm 0.04$ Ga and $Z_0 = 0.0155 \pm
0.0005$,  with present-day surface abundances 
$Y_{\rm s} = 0.224$, $Z_{\rm s} = 0.0142$.
It is interesting that the value found for 
$Z_{\rm s}$ is closer to modern spectroscopically determined values -- 
$0.0134$ \citep{Asplundetal2009ARA&A} and $0.0153$ \citep{Caffauetal2011SolPh}  -- than the value  $0.018$ of Model S whose
sound speed in the radiative envelope is more-or-less correct.

Recently \citet{doganboannoJCDsurfaceeffectsandage2010} have
refined the single-parameter calibration, obtaining 
$t_\odot = 4.57 \pm 0.08$ Ga, with an imposed  heavy-element abundance adjusted
to yield $Z_{\rm s} / X_{\rm s} = 0.0245$, as in Model S.  This age is
essentially in agreement with Bonanno, Schattl and Patern\`o's earlier result.
Meteoritic ages lie between $4.563$ and $4.576$ Ga  \citep{amelinmeteoriticages2002Sci, Jacobsenetalmeteoriticages2008E&PSL, Jacobsenetalerratum2009E&PSL, bouvieretalmeteriticages2010NatGe}. On considering the 
the two-parameter calibration of \citet{GHDOGsolarage2011MNRAS}, Christensen-Dalsgaard
privately complained that the latter leads to a model that is
seismically unacceptable (as, of course, it must be, because the
opacity is too low, and therefore so too is the sound speed in the
radiative envelope outside the core).  That 
remark begs the interesting question, to which I have already alluded,  
of whether or not with an
imperfect model it is better to satisfy conditions in the envelope
when trying to assess conditions in the core.  This is an important
issue of principle for model calibration -- I invite you to ignore the
fact that in this particular example the final answers agree to within
their quoted precision, for that is beside the point -- to which I now turn my attention.  

\section{The effect of hidden parameters on model-fitting}
With the solar age calibration in mind, I consider a class of solar models depending explicitly on two parameters, $\xi$ and $\zeta$, which could be $t_\odot$ and $Z_0$. Bearing in mind that none of those models is seismically consistent with the sun, I imagine there to be a broader, virtual,  class, extended in some (unknown to me) way so as to encompass the sun, and represent the distance from the sun of any of my explicit models to be characterized by the (hidden) parameter $\eta$, whose value is unknowable to the calibration.  I normalize the parameters in such a way that ($\xi$, $\eta$, $\zeta$)= (0, 0, 0) represents the real sun, and I presume that the largest values of ($\xi$, $\eta$, $\zeta$) that I need to consider are small enough to permit linearization of the model differences from the sun. I point out that in real life $\xi$, $\eta$, $\zeta$ are likely to be functions (or simply vectors, if those functions are considered to have been expanded in terms of a basis set).

I now consider making a series of different observations $O_i$, $i=1,2,...I,$ subject to random unknown errors $\epsilon_i$ which for simplicity I take to be independent.   The  values of $O_i$ are functionals of the structure. They too are normalized such that their exact values for the sun are zero. They can therefore be represented as a linear combination of the model-specifying parameters thus:
\begin{equation}
\label{hid3}
O_i = \alpha_i \xi + \beta_i \eta + \gamma_i \zeta - \epsilon_i = 0\, .
\end{equation}
The intention is to calibrate the models with these data, the principal interest being in the value of $\xi$.

The coefficients $\alpha_i$ and $\gamma_i$ are, of course, known functionals of the models, but we have no idea of the relation between $\beta_i$ and the models, nor even the measurements. Nevertheless, I assume for simplicity that the parameters ($\alpha_i, \beta_i, \gamma_i$) are scaled such that the data errors $\epsilon_i$ have the same variance $\sigma^2$. Then it is reasonable to attempt a calibration by minimizing
\begin{equation}
\label{hid4}
{\cal E} := \sum_{i=1}^I (\alpha_i \xi + \beta_i \eta + \gamma_i \zeta - \epsilon_i)^2 \, .
\end{equation}
The results will be found to depend on the known coefficients
\begin{equation}
\label{hid5}
A = \sum_i  \! \alpha_i ^2, \,\,\,\, B= \sum_i  \!\gamma_i\alpha_i,\,\,\,\, C=\sum_i \!\gamma_i ^2 
\end{equation}
and the error combinations
\begin{equation}
\label{hid6}
\epsilon_a=\sum_i\!\alpha_i\epsilon_i,\,\,\,\, \epsilon_b=\sum_i\!\gamma_i\epsilon_i\, ;
\end{equation}
they depend also on the unknown coefficients
\begin{equation}
\label{hid7}
D=\sum_i\!\alpha_i\beta_i,\,\,\,\,E=\sum_i\!\beta_i\gamma_i\, .
\end{equation}
\subsection{Full calibration}
The full calibration is the result of minimizing $\cal E$ with respect to both $\xi$ and $\eta$:
\begin{equation}
\label{hid8}
\xi=\xi_{\rm f}:=\Delta^{-1}[(BE-CD)\eta+B\epsilon_b-C\epsilon_a]\,;\,\,\,\,\,\,\Delta=AC-B^2\, .
\end{equation}
It has expectation
\begin{equation}
\label{hid9}
{\overline \xi_{\rm f}}=\Delta^{-1}(BE-CD)\eta
\end{equation}
and error-variance
\begin{equation}
\label{hid10}
\overline {\epsilon^2_{\xi_{\rm f}}}=\Delta^{-1}C\sigma^2=:\sigma_{\rm f}^2\, .
\end{equation}
The corresponding expressions for  $\zeta_{\rm f}$ are similar.  Note that the error variance is a measure of the precision of the calibration procedure; it is not an absolute indicator of the accuracy of the outcome.
\subsection{Partial calibration}
Minimizing $\cal E$ with respect to $\xi$ for an assumed value of $\zeta$ yields
\begin{equation}
\label{hid11}
\xi=\xi_{\rm p}:=-A^{-1}(B\zeta+D\eta+\epsilon_a) \, ,
\end{equation}
yielding
\begin{equation}
\label{hid12}
\overline{\xi_{\rm p}}=-A^{-1}(B\zeta+D\eta)\,,\,\,\,\,\overline{\epsilon^2_{\xi_{\rm p}}}=A^{-1}\sigma^2=:\sigma_{\rm p}^2\, .
\end{equation}
It is perhaps pertinent to remark that as the number $I$ of measurements is increased, the error-variance 
$\sigma_{\rm p}^2$ decreases, because $A$ increases: adding new information, taken properly into account, never decreases precision. That is true also when the errors are correlated.  
\subsection{Comparison of the calibrations}
We first notice that had the models encompassed the sun the parameter $\eta$ would have been redundant. It could have been set to zero, which is equivalent to having ignored the possibility of its existence. Then $\overline{\xi_{\rm f}} = 0$, the correct answer; but $\overline{\xi_{\rm p}} = - A^{-1}B \zeta$, which is unlikely to vanish unless the correct value of  $\zeta$, namely zero, were adopted. However, that is not the situation for most calibrations, entire solar models being an archetypical example.

In the pertinent case $\eta$ is not zero, and neither $\overline{\xi_{\rm f}}$ nor $\overline{\xi_{\rm p}}$ is likely to vanish.
Moreover, because each is susceptible to the unknown $\beta_i$ in different ways, one cannot compare the offsets in a reliable way, even when $\zeta = 0$ in the partial calibration. However, one can readily show, from Schwarz's inequality, that $\Delta > 0$, and hence
\begin{equation}
\label{hid13}
\sigma_{\rm f}^2-\sigma_{\rm p}^2=\Delta^{-1}B^2A^{-1} > 0\, ;
\end{equation}
reducing the parameter space of the calibration necessarily increases the precision (whether or not the measurement errors are independent). But does it improve the accuracy, even when the correct choice of the fixed parameters is made?

The $\eta$-dependent offsets given by equations (\ref{hid9}) and (\ref{hid12}) cannot be compared because we have no idea of the values of the coefficients $D$ and $E$. We don't even know physically what they mean, for they represent what we don't know. But as an exercise with possibly very little meaning -- indeed it is certainly not easy to interpret the results -- I have conducted statistical computations in which the coefficients ($\alpha_i, \beta_i, \gamma_i)$ and $\eta$) were varied randomly  (initially each uniformly and independently distributed between $-1$  and $+1$) to simulate random situations under calibration, in each case drawing a sufficient number of examples to obviate serious stochasticity in the outcome: 100 averages of 1000 examples with differing $\eta$ of using $I = 1000$ different measurements defined by  $(\alpha_i, \beta_i, \gamma_i)$ in a calibration. In the first case I assumed ignorance of $\zeta$, which in the partial calibration I drew  from a set of random numbers distributed identically to the other parameters. Both calibrations yielded small average values $\overline{\xi_{\rm f}^2}$ and $\overline {\xi_{\rm p}^2}$ -- they should have, because $\beta_i \eta$ were distributed symmetrically about zero -- although in 70 per cent of the cases  $\overline{\xi_{\rm f}^2}$ was smaller. When instead $\zeta$ was set to zero -- the correct value -- in about 70 per cent of the partial calibrations  $\overline {\xi_{\rm p}^2}$ was smaller. However, when $\zeta$ is set  to an incorrect value, significantly different from zero, then $\overline{\xi_{\rm f}^2}$ was almost always smaller.   Qualitatively similar results were obtained when 
$(\alpha_i,\beta_i,\gamma_i)$ and $\eta$ were distributed normally.   I accept that perusal of equations (\ref{hid9})-(\ref{hid12}) 
shows that the outcome must depend on the details of the distribution functions chosen for the parameters ($\alpha_i, \beta_i, \gamma_i; \eta$) that define the situations being calibrated.  Nonetheless, if one simply accepts the two simple distributions that I have adopted, then taking all these results together suggests the following rule of thumb: 
\begin{displaymath}
\label{hid14}
{\rm Accuracy\,\, declines\,\, with\,\, increasing \,\,precision.}
\end{displaymath}
\noindent It is extremely important to appreciate that this rule is merely a guide. 
It is not an inviolable law. Indeed, there are many obvious counterexamples. But it may serve the useful purpose of inducing one to investigate more carefully the analysis that one has in hand.

The rule serves the sobering purpose of suggesting that a partial calibration is likely to be less reliable than one incorporating a broader space of possibilities, even though the former is more precise. It is not uncommon, both in our discipline and elsewhere in science, for a careful extensively researched calibration to be repeated by a more restricted one, necessarily with higher precision, and the latter subsequently being used by others in preference in the possibly mistaken belief that it is more accurate.

I conclude with a few more obvious remarks in this vein. First it goes without saying that precision improves with better data. And accuracy does not decline -- indeed, it must improve too. But better data are not required for the direct purpose in hand if uncertainty in the calibration is dominated by $\eta$. Of course, it is not always evident how such a situation might be detected, let alone assessed. Furthermore, better data might also be useful indirectly, when using different procedures,  for assessing the deleterious impact of $\eta$. There are also obvious classes of procedures that can increase accuracy by decreasing parameter space, perhaps even more than the amount by which precision is increased. An obvious example is the application of constraints on the manner in which the data $O_i$ are combined in the calibration to eliminate contamination by known aspects of the model whose actual influence on $O_i$ is not known, analogous to the elimination of near surface effects in structure inversions via the function ${\cal P}/I$ in equations (\ref{delomegaintegrals}) and (\ref{inversionresiduals}).

\section{Summary}
Simple models have been useful for providing an initial rough understanding of the structure
of the sun and the manner in which the properties of helioseismic oscillations can be used for structural and kinematical diagnosis.  The analyses can be of toy models or asymptotic approximations to the real sun.  The analytical formula so obtained have played an extremely important role in designing diagnostic procedures.  It has been argued that numerical surveys can be just as good, and that the simple analytical procedures are therefore unnecessary.  I do not deny the sentiment behind that opinion.  But I do deny the reality. To be sure, in the right hands numerical surveys have contributed greatly to our knowledge: partly directly, and partly by providing benchmarks against which the analytical formulae can be tested and sometimes calibrated.  But it is also important to realize that in practice it has been principally the approximate analytical formulae that have motivated the design of the seismic diagnostics in use today; those formulae may not have been necessary, but they have surely accelerated the development of our ability to advance understanding.  To quote a few examples, it was simple models that led to the first seismic calibration of the depth of the solar convection zone, and the determination of the helium abundance, the interior rotation and some aspects of the structure of the core.  It has also been useful to adopt hybrid procedures, using, for example, simple, sometimes asymptotic, methods to estimate deviations of numerically computed models from the real sun.  Initial investigations of these matters have been followed up by other procedures, necessary not only to improve precision but, more importantly, to view the situation differently in order to detect whether particular individual procedures are  biassed by hidden agents.  By so doing, the reliability of the inferences is increased, sometimes causing estimates of accuracy to be moderated.  This point is illustrated by contrasting the first direct seismological determination of the depth of the convection zone, using a small suite of methods whose differences determined the accuracy, with subsequent single-procedure determinations with necessarily greater precision to which have  too hastily been 
attributed concomitant accuracy.  The simple toy model discussed \S 6 exemplifies that point.    

The broad message that I am trying to put forward is that helioseismology is not dead. Although for most astronomers asteroseismology offers a wider arena of discovery, for the physicist there is still much to investigate in the sun, possibly more than in other stars, at least in the short term when there is yet too much that is unknown about other stars to isolate issues in physics from uncertainties in stellar structure. I admit there have been counterexamples. Therefore, as I have illustrated in this discussion, there remains a substantial amount of work that theorists must attend to. There is also more work for data analysts too, who, sadly, are few. I repeat my plea for more attention to estimating error correlations, for they can influence inferences significantly \citep[e.g.][]{howethompsonerrorcorrelation1996MNRAS, DOGTakashierrorcorrelation2002MNRAS}, possibly even to the extent of biassing results  by an amount that is much greater than the apparent variance of the propagated random uncertainty \citep{DOGinverseproblemTenerifewinterschool1996stsu}. The community investigating such matters is small. But the importance of the endeavour is not.

\section{Acknowledgements}
\noindent
I thank Guenter Houdek and Masao Takata for stimulating discussions, and Guenter again for preparing Figures 1 and 2, and Holly Pearce for typing the draft manuscript.  
I am grateful to the Leverhulme Trust for an Emeritus Fellowship.
\vspace{-12pt}

\bibliography{../references}

\begin{thebibliography}{}
\expandafter\ifx\csname natexlab\endcsname\relax\def\natexlab#1{#1}\fi
\expandafter\ifx\csname url\endcsname\relax
  \def\url#1{\texttt{#1}}\fi
\expandafter\ifx\csname urlprefix\endcsname\relax\def\urlprefix{URL }\fi
\providecommand{\eprint}[2][]{\url{#2}}

\bibitem[{{Ahmad} et~al.(2002){Ahmad}, {Allen}, {Andersen}, {Anglin}, {Barton},
  {Beier}, {Bercovitch}, {Bigu}, {Biller}, {Black}, {Blevis}, {Boardman},
  {Boger}, {Bonvin}, {Boulay}, {Bowler}, {Bowles}, {Brice}, {Browne},
  {Bullard}, {B{\"u}hler}, {Cameron}, {Chan}, {Chen}, {Chen}, {Chen},
  {Cleveland}, {Clifford}, {Cowan}, {Cowen}, {Cox}, {Dai}, {Dalnoki-Veress},
  {Davidson}, {Doe}, {Doucas}, {Dragowsky}, {Duba}, {Duncan}, {Dunford},
  {Dunmore}, {Earle}, {Elliott}, {Evans}, {Ewan}, {Farine}, {Fergani},
  {Ferraris}, {Ford}, {Formaggio}, {Fowler}, {Frame}, {Frank}, {Frati},
  {Gagnon}, {Germani}, {Gil}, {Graham}, {Grant}, {Hahn}, {Hallin}, {Hallman},
  {Hamer}, {Hamian}, {Handler}, {Haq}, {Hargrove}, {Harvey}, {Hazama},
  {Heeger}, {Heintzelman}, {Heise}, {Helmer}, {Hepburn}, {Heron}, {Hewett},
  {Hime}, {Howe}, {Hykawy}, {Isaac}, {Jagam}, {Jelley}, {Jillings}, {Jonkmans},
  {Kazkaz}, {Keener}, {Klein}, {Knox}, {Komar}, {Kouzes}, {Kutter}, {Kyba},
  {Law}, {Lawson}, {Lay}, {Lee}, {Lesko}, {Leslie}, {Levine}, {Locke}, {Luoma},
  {Lyon}, {Majerus}, {Mak}, {Maneira}, {Manor}, {Marino}, {McCauley},
  {McDonald}, {McDonald}, {McFarlane}, {McGregor}, {Meijer Drees}, {Mifflin},
  {Miller}, {Milton}, {Moffat}, {Moorhead}, {Nally}, {Neubauer}, {Newcomer},
  {Ng}, {Noble}, {Norman}, {Novikov}, {O'Neill}, {Okada}, {Ollerhead}, {Omori},
  {Orrell}, {Oser}, {Poon}, {Radcliffe}, {Roberge}, {Robertson}, {Robertson},
  {Rosendahl}, {Rowley}, {Rusu}, {Saettler}, {Schaffer}, {Schwendener},
  {Sch{\"u}lke}, {Seifert}, {Shatkay}, {Simpson}, {Sims}, {Sinclair},
  {Skensved}, {Smith}, {Smith}, {Spreitzer}, {Starinsky}, {Steiger},
  {Stokstad}, {Stonehill}, {Storey}, {Sur}, {Tafirout}, {Tagg}, {Tanner},
  {Taplin}, {Thorman}, {Thornewell}, {Trent}, {Tserkovnyak}, {van Berg}, {van
  de Water}, {Virtue}, {Waltham}, {Wang}, {Wark}, {West}, {Wilhelmy},
  {Wilkerson}, {Wilson}, {Wittich}, {Wouters}, \&
  {Yeh}}]{ahmadetal_a_2002PhRvL..89a1301A}
{Ahmad}, Q.~R., {Allen}, R.~C., {Andersen}, T.~C., {Anglin}, J.~D., {Barton},
  J.~C., {Beier}, E.~W., {Bercovitch}, M., {Bigu}, J., {Biller}, S.~D.,
  {Black}, R.~A., {Blevis}, I., {Boardman}, R.~J., {Boger}, J., {Bonvin}, E.,
  {Boulay}, M.~G., {Bowler}, M.~G., {Bowles}, T.~J., {Brice}, S.~J., {Browne},
  M.~C., {Bullard}, T.~V., {B{\"u}hler}, G., {Cameron}, J., {Chan}, Y.~D.,
  {Chen}, H.~H., {Chen}, M., {Chen}, X., {Cleveland}, B.~T., {Clifford}, E.~T.,
  {Cowan}, J.~H., {Cowen}, D.~F., {Cox}, G.~A., {Dai}, X., {Dalnoki-Veress},
  F., {Davidson}, W.~F., {Doe}, P.~J., {Doucas}, G., {Dragowsky}, M.~R.,
  {Duba}, C.~A., {Duncan}, F.~A., {Dunford}, M., {Dunmore}, J.~A., {Earle},
  E.~D., {Elliott}, S.~R., {Evans}, H.~C., {Ewan}, G.~T., {Farine}, J.,
  {Fergani}, H., {Ferraris}, A.~P., {Ford}, R.~J., {Formaggio}, J.~A.,
  {Fowler}, M.~M., {Frame}, K., {Frank}, E.~D., {Frati}, W., {Gagnon}, N.,
  {Germani}, J.~V., {Gil}, S., {Graham}, K., {Grant}, D.~R., {Hahn}, R.~L.,
  {Hallin}, A.~L., {Hallman}, E.~D., {Hamer}, A.~S., {Hamian}, A.~A.,
  {Handler}, W.~B., {Haq}, R.~U., {Hargrove}, C.~K., {Harvey}, P.~J., {Hazama},
  R., {Heeger}, K.~M., {Heintzelman}, W.~J., {Heise}, J., {Helmer}, R.~L.,
  {Hepburn}, J.~D., {Heron}, H., {Hewett}, J., {Hime}, A., {Howe}, M.,
  {Hykawy}, J.~G., {Isaac}, M.~C., {Jagam}, P., {Jelley}, N.~A., {Jillings},
  C., {Jonkmans}, G., {Kazkaz}, K., {Keener}, P.~T., {Klein}, J.~R., {Knox},
  A.~B., {Komar}, R.~J., {Kouzes}, R., {Kutter}, T., {Kyba}, C.~C., {Law}, J.,
  {Lawson}, I.~T., {Lay}, M., {Lee}, H.~W., {Lesko}, K.~T., {Leslie}, J.~R.,
  {Levine}, I., {Locke}, W., {Luoma}, S., {Lyon}, J., {Majerus}, S., {Mak},
  H.~B., {Maneira}, J., {Manor}, J., {Marino}, A.~D., {McCauley}, N.,
  {McDonald}, A.~B., {McDonald}, D.~S., {McFarlane}, K., {McGregor}, G.,
  {Meijer Drees}, R., {Mifflin}, C., {Miller}, G.~G., {Milton}, G., {Moffat},
  B.~A., {Moorhead}, M., {Nally}, C.~W., {Neubauer}, M.~S., {Newcomer}, F.~M.,
  {Ng}, H.~S., {Noble}, A.~J., {Norman}, E.~B., {Novikov}, V.~M., {O'Neill},
  M., {Okada}, C.~E., {Ollerhead}, R.~W., {Omori}, M., {Orrell}, J.~L., {Oser},
  S.~M., {Poon}, A.~W., {Radcliffe}, T.~J., {Roberge}, A., {Robertson}, B.~C.,
  {Robertson}, R.~G., {Rosendahl}, S.~S., {Rowley}, J.~K., {Rusu}, V.~L.,
  {Saettler}, E., {Schaffer}, K.~K., {Schwendener}, M.~H., {Sch{\"u}lke}, A.,
  {Seifert}, H., {Shatkay}, M., {Simpson}, J.~J., {Sims}, C.~J., {Sinclair},
  D., {Skensved}, P., {Smith}, A.~R., {Smith}, M.~W., {Spreitzer}, T.,
  {Starinsky}, N., {Steiger}, T.~D., {Stokstad}, R.~G., {Stonehill}, L.~C.,
  {Storey}, R.~S., {Sur}, B., {Tafirout}, R., {Tagg}, N., {Tanner}, N.~W.,
  {Taplin}, R.~K., {Thorman}, M., {Thornewell}, P.~M., {Trent}, P.~T.,
  {Tserkovnyak}, Y.~I., {van Berg}, R., {van de Water}, R.~G., {Virtue}, C.~J.,
  {Waltham}, C.~E., {Wang}, J.-X., {Wark}, D.~L., {West}, N., {Wilhelmy},
  J.~B., {Wilkerson}, J.~F., {Wilson}, J.~R., {Wittich}, P., {Wouters}, J.~M.,
  \& {Yeh}, M. 2002, Physical Review Letters, 89, 011301.
  \eprint{arXiv:nucl-ex/0204008}

\bibitem[{{Ahmad} et~al.(2001){Ahmad}, {Allen}, {Andersen}, {Anglin},
  {B{\"u}hler}, {Barton}, {Beier}, {Bercovitch}, {Bigu}, {Biller}, {Black},
  {Blevis}, {Boardman}, {Boger}, {Bonvin}, {Boulay}, {Bowler}, {Bowles},
  {Brice}, {Browne}, {Bullard}, {Burritt}, {Cameron}, {Cameron}, {Chan},
  {Chen}, {Chen}, {Chen}, {Chon}, {Cleveland}, {Clifford}, {Cowan}, {Cowen},
  {Cox}, {Dai}, {Dai}, {Dalnoki-Veress}, {Davidson}, {Doe}, {Doucas},
  {Dragowsky}, {Duba}, {Duncan}, {Dunmore}, {Earle}, {Elliott}, {Evans},
  {Ewan}, {Farine}, {Fergani}, {Ferraris}, {Ford}, {Fowler}, {Frame}, {Frank},
  {Frati}, {Germani}, {Gil}, {Goldschmidt}, {Grant}, {Hahn}, {Hallin},
  {Hallman}, {Hamer}, {Hamian}, {Haq}, {Hargrove}, {Harvey}, {Hazama},
  {Heaton}, {Heeger}, {Heintzelman}, {Heise}, {Helmer}, {Hepburn}, {Heron},
  {Hewett}, {Hime}, {Howe}, {Hykawy}, {Isaac}, {Jagam}, {Jelley}, {Jillings},
  {Jonkmans}, {Karn}, {Keener}, {Kirch}, {Klein}, {Knox}, {Komar}, {Kouzes},
  {Kutter}, {Kyba}, {Law}, {Lawson}, {Lay}, {Lee}, {Lesko}, {Leslie}, {Levine},
  {Locke}, {Lowry}, {Luoma}, {Lyon}, {Majerus}, {Mak}, {Marino}, {McCauley},
  {McDonald}, {McDonald}, {McFarlane}, {McGregor}, {McLatchie}, {Drees}, {Mes},
  {Mifflin}, {Miller}, {Milton}, {Moffat}, {Moorhead}, {Nally}, {Neubauer},
  {Newcomer}, {Ng}, {Noble}, {Norman}, {Novikov}, {O'Neill}, {Okada},
  {Ollerhead}, {Omori}, {Orrell}, {Oser}, {Poon}, {Radcliffe}, {Roberge},
  {Robertson}, {Robertson}, {Rowley}, {Rusu}, {Saettler}, {Schaffer},
  {Schuelke}, {Schwendener}, {Seifert}, {Shatkay}, {Simpson}, {Sinclair},
  {Skensved}, {Smith}, {Smith}, {Starinsky}, {Steiger}, {Stokstad}, {Storey},
  {Sur}, {Tafirout}, {Tagg}, {Tanner}, {Taplin}, {Thorman}, {Thornewell},
  {Trent}, {Tserkovnyak}, {van Berg}, {van de Water}, {Virtue}, {Waltham},
  {Wang}, {Wark}, {West}, {Wilhelmy}, {Wilkerson}, {Wilson}, {Wittich},
  {Wouters}, \& {Yeh}}]{ahmadetalneutrinos2001PhRvL..87g1301A}
{Ahmad}, Q.~R., {Allen}, R.~C., {Andersen}, T.~C., {Anglin}, J.~D.,
  {B{\"u}hler}, G., {Barton}, J.~C., {Beier}, E.~W., {Bercovitch}, M., {Bigu},
  J., {Biller}, S., {Black}, R.~A., {Blevis}, I., {Boardman}, R.~J., {Boger},
  J., {Bonvin}, E., {Boulay}, M.~G., {Bowler}, M.~G., {Bowles}, T.~J., {Brice},
  S.~J., {Browne}, M.~C., {Bullard}, T.~V., {Burritt}, T.~H., {Cameron}, K.,
  {Cameron}, J., {Chan}, Y.~D., {Chen}, M., {Chen}, H.~H., {Chen}, X., {Chon},
  M.~C., {Cleveland}, B.~T., {Clifford}, E.~T., {Cowan}, J.~H., {Cowen}, D.~F.,
  {Cox}, G.~A., {Dai}, Y., {Dai}, X., {Dalnoki-Veress}, F., {Davidson}, W.~F.,
  {Doe}, P.~J., {Doucas}, G., {Dragowsky}, M.~R., {Duba}, C.~A., {Duncan},
  F.~A., {Dunmore}, J., {Earle}, E.~D., {Elliott}, S.~R., {Evans}, H.~C.,
  {Ewan}, G.~T., {Farine}, J., {Fergani}, H., {Ferraris}, A.~P., {Ford}, R.~J.,
  {Fowler}, M.~M., {Frame}, K., {Frank}, E.~D., {Frati}, W., {Germani}, J.~V.,
  {Gil}, S., {Goldschmidt}, A., {Grant}, D.~R., {Hahn}, R.~L., {Hallin}, A.~L.,
  {Hallman}, E.~D., {Hamer}, A., {Hamian}, A.~A., {Haq}, R.~U., {Hargrove},
  C.~K., {Harvey}, P.~J., {Hazama}, R., {Heaton}, R., {Heeger}, K.~M.,
  {Heintzelman}, W.~J., {Heise}, J., {Helmer}, R.~L., {Hepburn}, J.~D.,
  {Heron}, H., {Hewett}, J., {Hime}, A., {Howe}, M., {Hykawy}, J.~G., {Isaac},
  M.~C., {Jagam}, P., {Jelley}, N.~A., {Jillings}, C., {Jonkmans}, G., {Karn},
  J., {Keener}, P.~T., {Kirch}, K., {Klein}, J.~R., {Knox}, A.~B., {Komar},
  R.~J., {Kouzes}, R., {Kutter}, T., {Kyba}, C.~C., {Law}, J., {Lawson}, I.~T.,
  {Lay}, M., {Lee}, H.~W., {Lesko}, K.~T., {Leslie}, J.~R., {Levine}, I.,
  {Locke}, W., {Lowry}, M.~M., {Luoma}, S., {Lyon}, J., {Majerus}, S., {Mak},
  H.~B., {Marino}, A.~D., {McCauley}, N., {McDonald}, A.~B., {McDonald}, D.~S.,
  {McFarlane}, K., {McGregor}, G., {McLatchie}, W., {Drees}, R.~M., {Mes}, H.,
  {Mifflin}, C., {Miller}, G.~G., {Milton}, G., {Moffat}, B.~A., {Moorhead},
  M., {Nally}, C.~W., {Neubauer}, M.~S., {Newcomer}, F.~M., {Ng}, H.~S.,
  {Noble}, A.~J., {Norman}, E.~B., {Novikov}, V.~M., {O'Neill}, M., {Okada},
  C.~E., {Ollerhead}, R.~W., {Omori}, M., {Orrell}, J.~L., {Oser}, S.~M.,
  {Poon}, A.~W., {Radcliffe}, T.~J., {Roberge}, A., {Robertson}, B.~C.,
  {Robertson}, R.~G., {Rowley}, J.~K., {Rusu}, V.~L., {Saettler}, E.,
  {Schaffer}, K.~K., {Schuelke}, A., {Schwendener}, M.~H., {Seifert}, H.,
  {Shatkay}, M., {Simpson}, J.~J., {Sinclair}, D., {Skensved}, P., {Smith},
  A.~R., {Smith}, M.~W., {Starinsky}, N., {Steiger}, T.~D., {Stokstad}, R.~G.,
  {Storey}, R.~S., {Sur}, B., {Tafirout}, R., {Tagg}, N., {Tanner}, N.~W.,
  {Taplin}, R.~K., {Thorman}, M., {Thornewell}, P., {Trent}, P.~T.,
  {Tserkovnyak}, Y.~I., {van Berg}, R., {van de Water}, R.~G., {Virtue}, C.~J.,
  {Waltham}, C.~E., {Wang}, J.-X., {Wark}, D.~L., {West}, N., {Wilhelmy},
  J.~B., {Wilkerson}, J.~F., {Wilson}, J., {Wittich}, P., {Wouters}, J.~M., \&
  {Yeh}, M. 2001, Physical Review Letters, 87, 071301.
  \eprint{arXiv:nucl-ex/0106015}

\bibitem[{{Amelin} et~al.(2002){Amelin}, {Krot}, {Hutcheon}, \&
  {Ulyanov}}]{amelinmeteoriticages2002Sci}
{Amelin}, Y., {Krot}, A.~N., {Hutcheon}, I.~D., \& {Ulyanov}, A.~A. 2002,
  Science, 297, 1678

\bibitem[{{Ando} \& {Osaki}(1975)}]{AndoOsaki1975PASJ}
{Ando}, H., \& {Osaki}, Y. 1975, \pasj, 27, 581

\bibitem[{{Antia} \& {Basu}(2006)}]{antiabasuZdetermination2006ApJ}
{Antia}, H.~M., \& {Basu}, S. 2006, \apj, 644, 1292.
  \eprint{arXiv:astro-ph/0603001}

\bibitem[{{Antia} et~al.(2008){Antia}, {Chitre}, \&
  {Gough}}]{antiachitregoughsolarKE2008A&A}
{Antia}, H.~M., {Chitre}, S.~M., \& {Gough}, D.~O. 2008, \aap, 477, 657.
  \eprint{0711.0799}

\bibitem[{{Asplund} et~al.(2005){Asplund}, {Grevesse}, \&
  {Sauval}}]{Asplundetal2005}
{Asplund}, M., {Grevesse}, N., \& {Sauval}, A.~J. 2005, in Cosmic Abundances as
  Records of Stellar Evolution and Nucleosynthesis, edited by {T.~G.~Barnes III
  \& F.~N.~Bash}, vol. 336 of Astronomical Society of the Pacific Conference
  Series, 25

\bibitem[{{Asplund} et~al.(2009){Asplund}, {Grevesse}, {Sauval}, \&
  {Scott}}]{Asplundetal2009ARA&A}
{Asplund}, M., {Grevesse}, N., {Sauval}, A.~J., \& {Scott}, P. 2009, \araa, 47,
  481. \eprint{0909.0948}

\bibitem[{{Bahcall}(1964)}]{bahcallsolarneutrinoprediction1964PhRvL}
{Bahcall}, J.~N. 1964, Physical Review Letters, 12, 300

\bibitem[{{Bahcall}(1966)}]{Bahcall15snu1966PhRvL}
--- 1966, Physical Review Letters, 17, 398

\bibitem[{{Bahcall}(2001)}]{bahcalltriumphforstellarevolution2001Natur}
--- 2001, \nat, 412, 29

\bibitem[{{Bahcall} et~al.(1968){Bahcall}, {Bahcall}, \&
  {Shaviv}}]{Bahcall2shaviv1968PhRvL}
{Bahcall}, J.~N., {Bahcall}, N.~A., \& {Shaviv}, G. 1968, Physical Review
  Letters, 20, 1209

\bibitem[{{Bahcall} et~al.(1998){Bahcall}, {Basu}, \&
  {Pinsonneault}}]{bahcallbasupinsonneault1998PhLB}
{Bahcall}, J.~N., {Basu}, S., \& {Pinsonneault}, M.~H. 1998, Physics Letters B,
  433, 1. \eprint{arXiv:astro-ph/9805135}

\bibitem[{{Bahcall} \& {Pinsonneault}(2000)}]{BP2000standardsolarmodels}
{Bahcall}, J.~N., \& {Pinsonneault}, M.~H. 2000, {URL},
  http://www.sns.ias.edu/~jnb

\bibitem[{{Bahcall} et~al.(2001){Bahcall}, {Pinsonneault}, \&
  {Basu}}]{Bahcallpinsonneaultbasu2001ApJ}
{Bahcall}, J.~N., {Pinsonneault}, M.~H., \& {Basu}, S. 2001, \apj, 555, 990.
  \eprint{arXiv:astro-ph/0010346}

\bibitem[{{Basu} \& {Antia}(2008)}]{basuantiaabundancereview2008PhR}
{Basu}, S., \& {Antia}, H.~M. 2008, Phys. Rep., 457, 217. \eprint{0711.4590}

\bibitem[{{Basu} \& {Christensen-Dalsgaard}(1997)}]{basujcdintrinsiceos1997A&A}
{Basu}, S., \& {Christensen-Dalsgaard}, J. 1997, \aap, 322, L5.
  \eprint{arXiv:astro-ph/9702162}

\bibitem[{{Basu} et~al.(1997){Basu}, {Christensen-Dalsgaard}, {Chaplin},
  {Elsworth}, {Isaak}, {New}, {Schou}, {Thompson}, \&
  {Tomczyk}}]{Basuetalbestfrequencies1997MNRAS}
{Basu}, S., {Christensen-Dalsgaard}, J., {Chaplin}, W.~J., {Elsworth}, Y.,
  {Isaak}, G.~R., {New}, R., {Schou}, J., {Thompson}, M.~J., \& {Tomczyk}, S.
  1997, \mnras, 292, 243. \eprint{arXiv:astro-ph/9702105}

\bibitem[{{Basu} et~al.(1999){Basu}, {D{\"a}ppen}, \&
  {Nayfonov}}]{basudappennayfonov1999ApJ}
{Basu}, S., {D{\"a}ppen}, W., \& {Nayfonov}, A. 1999, \apj, 518, 985.
  \eprint{arXiv:astro-ph/9810132}

\bibitem[{{Baturin} et~al.(2000){Baturin}, {D{\"a}ppen}, {Gough}, \&
  {Vorontsov}}]{BaturinWDDOGSVV2000MNRAS}
{Baturin}, V.~A., {D{\"a}ppen}, W., {Gough}, D.~O., \& {Vorontsov}, S.~V. 2000,
  \mnras, 316, 71

\bibitem[{{Bonanno} et~al.(2002){Bonanno}, {Schlattl}, \&
  {Patern{\`o}}}]{bonannoschattlpaternosolarage2002A&A}
{Bonanno}, A., {Schlattl}, H., \& {Patern{\`o}}, L. 2002, \aap, 390, 1115.
  \eprint{arXiv:astro-ph/0204331}

\bibitem[{{Bouvier} \& {Wadhwa}(2010)}]{bouvieretalmeteriticages2010NatGe}
{Bouvier}, A., \& {Wadhwa}, M. 2010, Nature Geoscience, 3, 637

\bibitem[{{Brans} \& {Dicke}(1961)}]{BransDicke1961}
{Brans}, C., \& {Dicke}, R.~H. 1961, Physical Review, 124, 925

\bibitem[{{Bretherton} \& {Spiegel}(1968)}]{brethertonspiegel1968ApJ}
{Bretherton}, F.~P., \& {Spiegel}, A.~E. 1968, \apjl, 153, L77

\bibitem[{{Brun} et~al.(1999){Brun}, {Turck-Chi{\`e}ze}, \&
  {Zahn}}]{brunt-czahnstandardsolarmodels1999ApJ}
{Brun}, A.~S., {Turck-Chi{\`e}ze}, S., \& {Zahn}, J.~P. 1999, \apj, 525, 1032.
  \eprint{arXiv:astro-ph/9906382}

\bibitem[{{Brun} et~al.(2000){Brun}, {Turck-Chi{\`e}ze}, \&
  {Zahn}}]{erratumtobrunt-czahnstandardsolarmodels2000ApJ}
--- 2000, \apj, 536, 1005

\bibitem[{{Brun} \& {Zahn}(2006)}]{brunzahn2006A&A}
{Brun}, A.~S., \& {Zahn}, J.-P. 2006, \aap, 457, 665.
  \eprint{arXiv:astro-ph/0610069}

\bibitem[{{Caffau} et~al.(2008){Caffau}, {Ludwig}, {Steffen}, {Ayres},
  {Bonifacio}, {Cayrel}, {Freytag}, \& {Plez}}]{caffauetal3dmodel2008A&A}
{Caffau}, E., {Ludwig}, H.-G., {Steffen}, M., {Ayres}, T.~R., {Bonifacio}, P.,
  {Cayrel}, R., {Freytag}, B., \& {Plez}, B. 2008, \aap, 488, 1031.
  \eprint{0805.4398}

\bibitem[{{Caffau} et~al.(2011){Caffau}, {Ludwig}, {Steffen}, {Freytag}, \&
  {Bonifacio}}]{Caffauetal2011SolPh}
{Caffau}, E., {Ludwig}, H.-G., {Steffen}, M., {Freytag}, B., \& {Bonifacio}, P.
  2011, \solphys, 268, 255. \eprint{1003.1190}

\bibitem[{{Caffau} et~al.(2009){Caffau}, {Maiorca}, {Bonifacio}, {Faraggiana},
  {Steffen}, {Ludwig}, {Kamp}, \& {Busso}}]{caffauetal2009A&A}
{Caffau}, E., {Maiorca}, E., {Bonifacio}, P., {Faraggiana}, R., {Steffen}, M.,
  {Ludwig}, H.-G., {Kamp}, I., \& {Busso}, M. 2009, \aap, 498, 877.
  \eprint{0903.3406}

\bibitem[{{Chaplin} et~al.(1996){Chaplin}, {Elsworth}, {Howe}, {Isaak},
  {McLeod}, {Miller}, {van der Raay}, {Wheeler}, \&
  {New}}]{chaplinetalfrequencies1996SoPh.168.1C}
{Chaplin}, W.~J., {Elsworth}, Y., {Howe}, R., {Isaak}, G.~R., {McLeod}, C.~P.,
  {Miller}, B.~A., {van der Raay}, H.~B., {Wheeler}, S.~J., \& {New}, R. 1996,
  \solphys, 168, 1

\bibitem[{{Charbonneau}(1998)}]{charbonneautachocline1998JRASC}
{Charbonneau}, P. 1998, \jrasc, 92, 311

\bibitem[{{Charbonneau} et~al.(1999){Charbonneau}, {Christensen-Dalsgaard},
  {Henning}, {Larsen}, {Schou}, {Thompson}, \&
  {Tomczyk}}]{charbonneauetaltachocline1999ApJ}
{Charbonneau}, P., {Christensen-Dalsgaard}, J., {Henning}, R., {Larsen}, R.~M.,
  {Schou}, J., {Thompson}, M.~J., \& {Tomczyk}, S. 1999, \apj, 527, 445

\bibitem[{{Christensen-Dalsgaard}(1988{\natexlab{a}})}]{JCDdiagram1988IAUS}
{Christensen-Dalsgaard}, J. 1988{\natexlab{a}}, in Advances in Helio- and
  Asteroseismology, edited by {J.~Christensen-Dalsgaard \& S.~Frandsen}, vol.
  123 of IAU Symposium, 295

\bibitem[{{Christensen-Dalsgaard}(1988{\natexlab{b}})}]{JCDhelioseismology1988}
--- 1988{\natexlab{b}}, in Seismology of the Sun and Sun-Like Stars, edited by
  {E.~J.~Rolfe}, vol. 286 of ESA Special Publication, 431

\bibitem[{{Christensen-Dalsgaard}(1991)}]{JCDoscillations1991}
--- 1991, in Challenges to Theories of the Structure of Moderate-Mass Stars,
  edited by {D.~O.~Gough \& J.~Toomre}, vol. 388 of Lecture Notes in Physics,
  Berlin Springer Verlag, 11

\bibitem[{{Christensen-Dalsgaard}(1996)}]{JCDtestingamodel1996}
--- 1996, in The Structure of the Sun, edited by {T.~Roca Cort{\'e}s \&
  F.~S{\'a}nchez}, 47

\bibitem[{{Christensen-Dalsgaard} \& {D{\"a}ppen}(1992)}]{JCDWDEoS1992A&ARv}
{Christensen-Dalsgaard}, J., \& {D{\"a}ppen}, W. 1992, \aapr, 4, 267

\bibitem[{{Christensen-Dalsgaard} et~al.(1996){Christensen-Dalsgaard},
  {D{\"a}ppen}, {Ajukov}, {Anderson}, {Antia}, {Basu}, {Baturin}, {Berthomieu},
  {Chaboyer}, {Chitre}, {Cox}, {Demarque}, {Donatowicz}, {Dziembowski},
  {Gabriel}, {Gough}, {Guenther}, {Guzik}, {Harvey}, {Hill}, {Houdek},
  {Iglesias}, {Kosovichev}, {Leibacher}, {Morel}, {Proffitt}, {Provost},
  {Reiter}, {Rhodes}, {Rogers}, {Roxburgh}, {Thompson}, \&
  {Ulrich}}]{jcdetal1996Sci}
{Christensen-Dalsgaard}, J., {D{\"a}ppen}, W., {Ajukov}, S.~V., {Anderson},
  E.~R., {Antia}, H.~M., {Basu}, S., {Baturin}, V.~A., {Berthomieu}, G.,
  {Chaboyer}, B., {Chitre}, S.~M., {Cox}, A.~N., {Demarque}, P., {Donatowicz},
  J., {Dziembowski}, W.~A., {Gabriel}, M., {Gough}, D.~O., {Guenther}, D.~B.,
  {Guzik}, J.~A., {Harvey}, J.~W., {Hill}, F., {Houdek}, G., {Iglesias}, C.~A.,
  {Kosovichev}, A.~G., {Leibacher}, J.~W., {Morel}, P., {Proffitt}, C.~R.,
  {Provost}, J., {Reiter}, J., {Rhodes}, E.~J., Jr, {Rogers}, F.~J.,
  {Roxburgh}, I.~W., {Thompson}, M.~J., \& {Ulrich}, R.~K. 1996, Science, 272,
  1286

\bibitem[{{Christensen-Dalsgaard} et~al.(2009){Christensen-Dalsgaard}, {Di
  Mauro}, {Houdek}, \& {Pijpers}}]{JCDetalopacity2009}
{Christensen-Dalsgaard}, J., {Di Mauro}, M.~P., {Houdek}, G., \& {Pijpers}, F.
  2009, \aap, 494, 205. \eprint{0811.1001}

\bibitem[{{Christensen-Dalsgaard} et~al.(1985){Christensen-Dalsgaard},
  {Duvall}, {Gough}, {Harvey}, \& {Rhodes}}]{speedofsoundJCDetal1985Natur}
{Christensen-Dalsgaard}, J., {Duvall}, T.~L., Jr, {Gough}, D.~O., {Harvey},
  J.~W., \& {Rhodes}, E.~J., Jr 1985, \nat, 315, 378

\bibitem[{{Christensen-Dalsgaard} \&
  {Gough}(1976)}]{jcddogheliologicalinverseproblem1976Natur}
{Christensen-Dalsgaard}, J., \& {Gough}, D.~O. 1976, \nat, 259, 89

\bibitem[{{Christensen-Dalsgaard} \& {Gough}(1981)}]{JCDDOG1981A&A}
--- 1981, \aap, 104, 173

\bibitem[{{D{\"a}ppen}(1998)}]{dappenEoS1998SSRv}
{D{\"a}ppen}, W. 1998, \ssr, 85, 49

\bibitem[{{D{\"a}ppen} et~al.(1990){D{\"a}ppen}, {Lebreton}, \&
  {Rogers}}]{dappenlebretonrogersEoS1990SoPh}
{D{\"a}ppen}, W., {Lebreton}, Y., \& {Rogers}, F. 1990, \solphys, 128, 35

\bibitem[{{D{\"a}ppen} \& {Nayfonov}(2000)}]{dappenanyfonovEoS2000ApJS}
{D{\"a}ppen}, W., \& {Nayfonov}, A. 2000, \apjs, 127, 287

\bibitem[{{Davis} et~al.(1968){Davis}, {Harmer}, \&
  {Hoffman}}]{davisharmerhoffmanneutrinos1968PhRvL}
{Davis}, J., R., {Harmer}, D.~S., \& {Hoffman}, K.~C. 1968, Physical Review
  Letters, 20, 1205

\bibitem[{{Deubner}(1975)}]{deubner1975A&A}
{Deubner}, F.-L. 1975, \aap, 44, 371

\bibitem[{{Di Mauro} et~al.(2002){Di Mauro}, {Christensen-Dalsgaard},
  {Rabello-Soares}, \& {Basu}}]{dimaurobasuetalintrinsicgamma_12002A&A}
{Di Mauro}, M.~P., {Christensen-Dalsgaard}, J., {Rabello-Soares}, M.~C., \&
  {Basu}, S. 2002, \aap, 384, 666

\bibitem[{{Dicke}(1964)}]{dicke1964Natur}
{Dicke}, R.~H. 1964, \nat, 202, 432

\bibitem[{{Dicke}(1967)}]{dickespindown1967ApJ}
--- 1967, \apjl, 149, L121

\bibitem[{{Dicke} \& {Goldenberg}(1967)}]{DickeGold1967PhRvL}
{Dicke}, R.~H., \& {Goldenberg}, H.~M. 1967, Physical Review Letters, 18, 313

\bibitem[{{Dicke} \& {Goldenberg}(1974)}]{dickegoldenberg1974ApJS}
--- 1974, \apjs, 27, 131

\bibitem[{{Do{\u g}an} et~al.(2010){Do{\u g}an}, {Bonanno}, \&
  {Christensen-Dalsgaard}}]{doganboannoJCDsurfaceeffectsandage2010}
{Do{\u g}an}, G., {Bonanno}, A., \& {Christensen-Dalsgaard}, J. 2010, ArXiv
  e-prints. \eprint{1004.2215}

\bibitem[{{Drake} \& {Testa}(2005)}]{draketesta2005Natur}
{Drake}, J.~J., \& {Testa}, P. 2005, \nat, 436, 525.
  \eprint{arXiv:astro-ph/0506182}

\bibitem[{{Duvall} et~al.(1984){Duvall}, {Dziembowski}, {Goode}, {Gough},
  {Harvey}, \& {Leibacher}}]{Naturerotation1984Natur}
{Duvall}, T.~L., Jr, {Dziembowski}, W.~A., {Goode}, P.~R., {Gough}, D.~O.,
  {Harvey}, J.~W., \& {Leibacher}, J.~W. 1984, \nat, 310, 22

\bibitem[{{Dziembowski} et~al.(1999){Dziembowski}, {Fiorentini}, {Ricci}, \&
  {Sienkiewicz}}]{dziembowskietalsolarage1999A&A}
{Dziembowski}, W.~A., {Fiorentini}, G., {Ricci}, B., \& {Sienkiewicz}, R. 1999,
  \aap, 343, 990. \eprint{arXiv:astro-ph/9809361}

\bibitem[{{Dziembowski} et~al.(1994){Dziembowski}, {Moskalik}, \&
  {Pamyatnykh}}]{DziembowskietalSPBstars1994IAUS}
{Dziembowski}, W.~A., {Moskalik}, P., \& {Pamyatnykh}, A.~A. 1994, in
  Pulsation; Rotation; and Mass Loss in Early-Type Stars, edited by
  {L.~A.~Balona, H.~F.~Henrichs, \& J.~M.~Le Contel}, vol. 162 of IAU
  Symposium, 69

\bibitem[{{Eddington}(1926)}]{eddingtoninternalconstitution1926ics}
{Eddington}, A.~S. 1926, {The Internal Constitution of the Stars: Cambridge
  University Press}

\bibitem[{{Einstein}(1926)}]{einsteinmeanders}
{Einstein}, A. 1926, Naturwissensch., 14

\bibitem[{{Elliott}(1996)}]{elliottseismiceos1996MNRAS}
{Elliott}, J.~R. 1996, \mnras, 280, 1244

\bibitem[{{Elliott}(1997)}]{elliott1997A&A}
--- 1997, \aap, 327, 1222

\bibitem[{{Elliott} et~al.(1998){Elliott}, {Gough}, \&
  {Sekii}}]{elliottdogsekiitachocline1998ESASP}
{Elliott}, J.~R., {Gough}, D.~O., \& {Sekii}, T. 1998, in Structure and
  Dynamics of the Interior of the Sun and Sun-like Stars, edited by
  {S.~Korzennik}, vol. 418 of ESA Special Publication, 763

\bibitem[{{Faulkner} \& {Swenson}(1988)}]{faulknerswenson1988ApJ}
{Faulkner}, J., \& {Swenson}, F.~J. 1988, \apjl, 329, L47

\bibitem[{{Garaud}(1999)}]{pascalebpropagationt1999MNRAS}
{Garaud}, P. 1999, \mnras, 304, 583

\bibitem[{{Garaud}(2002)}]{garaudtachoclineI2002MNRAS}
--- 2002, \mnras, 329, 1. \eprint{arXiv:astro-ph/0108276}

\bibitem[{{Garaud} \& {Acevedo Arreguin}(2009)}]{garaudacevedoarreguin2009ApJ}
{Garaud}, P., \& {Acevedo Arreguin}, L. 2009, \apj, 704, 1. \eprint{0906.1756}

\bibitem[{{Garaud} \& {Garaud}(2008)}]{garaudgaraud2008MNRAS}
{Garaud}, P., \& {Garaud}, J.-D. 2008, \mnras, 391, 1239. \eprint{0806.2551}

\bibitem[{{Gough}(1977)}]{DOG1977Nice}
{Gough}, D.~O. 1977, in IAU Colloq. 36: The Energy Balance and Hydrodynamics of
  the Solar Chromosphere and Corona, edited by {B.~Bonnet \& P.~Delache}, G. de
  Bussac, Clermont-Ferrand, 3

\bibitem[{{Gough}(1983{\natexlab{a}})}]{DOG1983PhysBull}
--- 1983{\natexlab{a}}, Physics Bulletin, 34, 502

\bibitem[{{Gough}(1983{\natexlab{b}})}]{DOGprotosolarY1983}
--- 1983{\natexlab{b}}, in Primordial Helium, edited by {P.~A.~Shaver,
  D.~Kunth, \& K.~Kjar}, 117

\bibitem[{{Gough}(1984)}]{DOGCatania1984MmSAI}
--- 1984, Mem. Soc. Astron. Ital., 55, 13

\bibitem[{{Gough}(1990{\natexlab{a}})}]{DOGOji1990LNP}
--- 1990{\natexlab{a}}, in Progress of Seismology of the Sun and Stars, edited
  by {Y.~Osaki \& H.~Shibahashi}, vol. 367 of Lecture Notes in Physics, Berlin
  Springer Verlag, 283

\bibitem[{{Gough}(1990{\natexlab{b}})}]{1990stromgren}
--- 1990{\natexlab{b}}, in Astrophysics: Recent Progress and Future
  Possibilities (A91-15054 03-90), Kongelige Danske Videnskabernes Selskab,
  Mat-fys Medd, {\bf 42}:4, edited by {B.~Gustafsson \& P.~E.~Nissen}, 13

\bibitem[{{Gough}(1993)}]{dog1993LH}
--- 1993, in Astrophysical Fluid Dynamics - Les Houches XLVII, 1987, edited by
  {J.-P.~Zahn \& J.~Zinn-Justin}, Elsevier, Amsterdam, 399

\bibitem[{{Gough}(1995)}]{DOGlinearXc-t1995ASPC}
--- 1995, in GONG 1994: Helio- and Astro-Seismology from the Earth and Space,
  edited by {R.~K.~Ulrich, E.~J.~Rhodes Jr, \& W.~D{\"a}ppen}, vol.~76 of
  Astronomical Society of the Pacific Conference Series, 551

\bibitem[{{Gough}(1996)}]{DOGinverseproblemTenerifewinterschool1996stsu}
--- 1996, in The Structure of the Sun, edited by {T.~Roca Cort{\'e}s \&
  F.~S{\'a}nchez}, 141

\bibitem[{{Gough}(2001)}]{solaragevhippel2001ASPC}
--- 2001, in Astrophysical Ages and Times Scales, edited by {T.~von Hippel,
  C.~Simpson, \& N.~Manset}, vol. 245 of Astronomical Society of the Pacific
  Conference Series, 31

\bibitem[{{Gough}(2003)}]{dog2003Ap&SS}
--- 2003, \apss, 285, 341

\bibitem[{{Gough}(2004)}]{DOGLorentz2004AIPC}
--- 2004, in Equation-of-State and Phase-Transition in Models of Ordinary
  Astrophysical Matter, edited by {V.~\v Celebonovi\'c, D.~Gough, \&
  W.~D{\"a}ppen}, vol. 731 of American Institute of Physics Conference Series,
  119

\bibitem[{{Gough}(2012)}]{DOG2012GApFD}
--- 2012, Geophysical and Astrophysical Fluid Dynamics, 106, 429

\bibitem[{{Gough} \& {McIntyre}(1998)}]{dogmem1998Nature}
{Gough}, D.~O., \& {McIntyre}, M.~E. 1998, \nat, 394, 755

\bibitem[{{Gough} \& {Novotny}(1990)}]{DOGnovotny1990SoPh}
{Gough}, D.~O., \& {Novotny}, E. 1990, \solphys, 128, 143

\bibitem[{{Gough} \& {Sekii}(2002)}]{DOGTakashierrorcorrelation2002MNRAS}
{Gough}, D.~O., \& {Sekii}, T. 2002, \mnras, 335, 170

\bibitem[{{Gough} \& {Vorontsov}(1995)}]{dogsvv1995MNRAS}
{Gough}, D.~O., \& {Vorontsov}, S.~V. 1995, \mnras, 273, 573

\bibitem[{{Gough} \& {Weiss}(1976)}]{DOGNOW1976MNRAS}
{Gough}, D.~O., \& {Weiss}, N.~O. 1976, \mnras, 176, 589

\bibitem[{{Grevesse} et~al.(2011){Grevesse}, {Asplund}, {Sauval}, \&
  {Scott}}]{GrevesseAsplundSauval2011sswh.book}
{Grevesse}, N., {Asplund}, M., {Sauval}, A.~J., \& {Scott}, P. 2011, in The
  Sun, the Solar Wind, and the Heliosphere, edited by {Miralles, M.~P.~\&
  S{\'a}nchez Almeida, J.}, 51

\bibitem[{{Grevesse} \& {Sauval}(1998)}]{grevessesauvalabundances1998SSRv}
{Grevesse}, N., \& {Sauval}, A.~J. 1998, \ssr, 85, 161

\bibitem[{{Guzik} \& {Mussack}(2010)}]{guzikmussack2010ApJ}
{Guzik}, J.~A., \& {Mussack}, K. 2010, \apj, 713, 1108

\bibitem[{{Guzik} et~al.(2005){Guzik}, {Watson}, \&
  {Cox}}]{guziketalsolarcontamination2005ApJ}
{Guzik}, J.~A., {Watson}, L.~S., \& {Cox}, A.~N. 2005, \apj, 627, 1049.
  \eprint{arXiv:astro-ph/0502364}

\bibitem[{{Houdek} \& {Gough}(2011)}]{GHDOGsolarage2011MNRAS}
{Houdek}, G., \& {Gough}, D.~O. 2011, \mnras, 418, 1217. \eprint{1108.0802}

\bibitem[{{Howard} et~al.(1967){Howard}, {Moore}, \&
  {Spiegel}}]{howardmoorespiegel67}
{Howard}, L.~N., {Moore}, D.~W., \& {Spiegel}, E.~A. 1967, Nature, 214, 1297

\bibitem[{{Howe} et~al.(2000){Howe}, {Christensen-Dalsgaard}, {Hill}, {Komm},
  {Larsen}, {Schou}, {Thompson}, \&
  {Toomre}}]{howeetaltachoclineoscillations2000Sci}
{Howe}, R., {Christensen-Dalsgaard}, J., {Hill}, F., {Komm}, R.~W., {Larsen},
  R.~M., {Schou}, J., {Thompson}, M.~J., \& {Toomre}, J. 2000, Science, 287,
  2456

\bibitem[{{Howe} et~al.(2011){Howe}, {Hill}, {Komm}, {Christensen-Dalsgaard},
  {Larson}, {Schou}, {Thompson}, \& {Ulrich}}]{howeetaltorsoscupdate2011JPhCS}
{Howe}, R., {Hill}, F., {Komm}, R., {Christensen-Dalsgaard}, J., {Larson},
  T.~P., {Schou}, J., {Thompson}, M.~J., \& {Ulrich}, R. 2011, Journal of
  Physics Conference Series, 271, 012074

\bibitem[{{Howe} \& {Thompson}(1996)}]{howethompsonerrorcorrelation1996MNRAS}
{Howe}, R., \& {Thompson}, M.~J. 1996, \mnras, 281, 1385

\bibitem[{{Hughes} et~al.(2007){Hughes}, {Rosner}, \&
  {Weiss}}]{hughesrosnerweisstach2007}
{Hughes}, D.~W., {Rosner}, R., \& {Weiss}, N.~O. (eds.) 2007, {The Solar
  Tachocline}

\bibitem[{{Iglesias} \& {Rogers}(1991)}]{Iglesiasrogersopacity1991ApJ}
{Iglesias}, C.~A., \& {Rogers}, F.~J. 1991, \apj, 371, 408

\bibitem[{{Iglesias} et~al.(1990){Iglesias}, {Rogers}, \&
  {Wilson}}]{iglesiasrogerscepheidopacity1990ApJ}
{Iglesias}, C.~A., {Rogers}, F.~J., \& {Wilson}, B.~G. 1990, \apj, 360, 221

\bibitem[{{Jacobsen} et~al.(2008){Jacobsen}, {Yin}, {Moynier}, {Amelin},
  {Krot}, {Nagashima}, {Hutcheon}, \&
  {Palme}}]{Jacobsenetalmeteoriticages2008E&PSL}
{Jacobsen}, B., {Yin}, Q.-Z., {Moynier}, F., {Amelin}, Y., {Krot}, A.~N.,
  {Nagashima}, K., {Hutcheon}, I.~D., \& {Palme}, H. 2008, Earth and Planetary
  Science Letters, 272, 353

\bibitem[{{Jacobsen} et~al.(2009){Jacobsen}, {Yin}, {Moynier}, {Amelin},
  {Krot}, {Nagashima}, {Hutcheon}, \& {Palme}}]{Jacobsenetalerratum2009E&PSL}
--- 2009, Earth and Planetary Science Letters, 277, 549

\bibitem[{{Jeffreys} \& {Swirles}(1956)}]{J2}
{Jeffreys}, H., \& {Swirles}, B. 1956, {Methods of Mathematical Physics:
  Cambridge Univ. Press}

\bibitem[{{Kosovichev}(1996)}]{AGKtachocline1996ApJ}
{Kosovichev}, A.~G. 1996, \apjl, 469, L61

\bibitem[{{Mel{\'e}ndez} et~al.(2009){Mel{\'e}ndez}, {Asplund}, {Gustafsson},
  \& {Yong}}]{solarlowZcontamination2009ApJ}
{Mel{\'e}ndez}, J., {Asplund}, M., {Gustafsson}, B., \& {Yong}, D. 2009, \apjl,
  704, L66. \eprint{0909.2299}

\bibitem[{{Moskalik} \&
  {Dziembowski}(1992)}]{moskalikdziembowskibetacephei1992A&A}
{Moskalik}, P., \& {Dziembowski}, W.~A. 1992, \aap, 256, L5

\bibitem[{{Mussack} \& {Gough}(2009)}]{kmdog2009}
{Mussack}, K., \& {Gough}, D.~O. 2009, in Solar-Stellar Dynamos as Revealed by
  Helio- and Asteroseismology: GONG 2008/SOHO 21, edited by {M.~Dikpati,
  T.~Arentoft, I.~Gonz{\'a}lez Hern{\'a}ndez, C.~Lindsey, \& F.~Hill}, vol. 416
  of Astronomical Society of the Pacific Conference Series, 203.
  \eprint{0810.2722}

\bibitem[{{Nordlund} et~al.(2009){Nordlund}, {Stein}, \&
  {Asplund}}]{nordlundsteinasplund2009LRSP}
{Nordlund}, {\AA}., {Stein}, R.~F., \& {Asplund}, M. 2009, Living Reviews in
  Solar Physics, 6, 2

\bibitem[{{Rabello-Soares} et~al.(2000){Rabello-Soares}, {Basu},
  {Christensen-Dalsgaard}, \& {Di
  Mauro}}]{rabellosoaresetalintrinsicgamma_12000SoPh}
{Rabello-Soares}, M.~C., {Basu}, S., {Christensen-Dalsgaard}, J., \& {Di
  Mauro}, M.~P. 2000, \solphys, 193, 345

\bibitem[{{Schou} et~al.(1998){Schou}, {Antia}, {Basu}, {Bogart}, {Bush},
  {Chitre}, {Christensen-Dalsgaard}, {di Mauro}, {Dziembowski}, {Eff-Darwich},
  {Gough}, {Haber}, {Hoeksema}, {Howe}, {Korzennik}, {Kosovichev}, {Larsen},
  {Pijpers}, {Scherrer}, {Sekii}, {Tarbell}, {Title}, {Thompson}, \&
  {Toomre}}]{schouetalrotation1998ApJ}
{Schou}, J., {Antia}, H.~M., {Basu}, S., {Bogart}, R.~S., {Bush}, R.~I.,
  {Chitre}, S.~M., {Christensen-Dalsgaard}, J., {di Mauro}, M.~P.,
  {Dziembowski}, W.~A., {Eff-Darwich}, A., {Gough}, D.~O., {Haber}, D.~A.,
  {Hoeksema}, J.~T., {Howe}, R., {Korzennik}, S.~G., {Kosovichev}, A.~G.,
  {Larsen}, R.~M., {Pijpers}, F.~P., {Scherrer}, P.~H., {Sekii}, T., {Tarbell},
  T.~D., {Title}, A.~M., {Thompson}, M.~J., \& {Toomre}, J. 1998, \apj, 505,
  390

\bibitem[{{Spiegel}(1972)}]{eastachycline1972NASSP}
{Spiegel}, E.~A. 1972, NASA Special Publication, 300, 61

\bibitem[{{Spiegel} \& {Zahn}(1992)}]{easjpztach1992A&A}
{Spiegel}, E.~A., \& {Zahn}, J.-P. 1992, \aap, 265, 106

\bibitem[{{Stein} \& {Nordlund}(1998)}]{steinnordlundatmosprops1998ApJ}
{Stein}, R.~F., \& {Nordlund}, A. 1998, \apj, 499, 914

\bibitem[{{Tomczyk} et~al.(1995{\natexlab{a}}){Tomczyk}, {Schou}, \&
  {Thompson}}]{tomczykschoumjt1995ApJ...448L..57T}
{Tomczyk}, S., {Schou}, J., \& {Thompson}, M.~J. 1995{\natexlab{a}}, \apjl,
  448, L57

\bibitem[{{Tomczyk} et~al.(1995{\natexlab{b}}){Tomczyk}, {Streander}, {Card},
  {Elmore}, {Hull}, \& {Cacciani}}]{tomczyketalLOWL1995SoPh..159....1T}
{Tomczyk}, S., {Streander}, K., {Card}, G., {Elmore}, D., {Hull}, H., \&
  {Cacciani}, A. 1995{\natexlab{b}}, \solphys, 159, 1

\bibitem[{{Trewin}(1967)}]{Trewin67}
{Trewin}, J.~C. 1967, {The Journal of William Charles Macready 1832--1851:
  Southern Illinois Univ. Press, Carbondale}

\bibitem[{{Tripathy} \&
  {Christensen-Dalsgaard}(1998)}]{tripahyjcdI_1998A&A...337..579T}
{Tripathy}, S.~C., \& {Christensen-Dalsgaard}, J. 1998, \aap, 337, 579.
  \eprint{arXiv:astro-ph/9709206}

\bibitem[{{Turck-Chi{\`e}ze} et~al.(2001){Turck-Chi{\`e}ze}, {Couvidat},
  {Kosovichev}, {Gabriel}, {Berthomieu}, {Brun}, {Christensen-Dalsgaard},
  {Garc{\'{\i}}a}, {Gough}, {Provost}, {Roca-Cortes}, {Roxburgh}, \&
  {Ulrich}}]{Turckchiezeetalseismicmodel2001ApJ}
{Turck-Chi{\`e}ze}, S., {Couvidat}, S., {Kosovichev}, A.~G., {Gabriel}, A.~H.,
  {Berthomieu}, G., {Brun}, A.~S., {Christensen-Dalsgaard}, J.,
  {Garc{\'{\i}}a}, R.~A., {Gough}, D.~O., {Provost}, J., {Roca-Cortes}, T.,
  {Roxburgh}, I.~W., \& {Ulrich}, R.~K. 2001, \apjl, 555, L69

\bibitem[{{Ulrich} \& {Rhodes}(1977)}]{rkuejrdepthconvzone1977ApJ}
{Ulrich}, R.~K., \& {Rhodes}, E.~J., Jr 1977, \apj, 218, 521

\bibitem[{{Vorontsov} et~al.(2002){Vorontsov}, {Christensen-Dalsgaard},
  {Schou}, {Strakhov}, \& {Thompson}}]{vorontsovetaltorsionalosc2002Sci}
{Vorontsov}, S.~V., {Christensen-Dalsgaard}, J., {Schou}, J., {Strakhov},
  V.~N., \& {Thompson}, M.~J. 2002, Science, 296, 101

\bibitem[{{Wood} et~al.(2011){Wood}, {McCaslin}, \&
  {Garaud}}]{woodmccaslingaraud2011ApJ}
{Wood}, T.~S., {McCaslin}, J.~O., \& {Garaud}, P. 2011, \apj, 738, 47.
  \eprint{1106.5250}

\bibitem[{{Wood} \& {McIntyre}(2011)}]{tswmemtach2011JFM}
{Wood}, T.~S., \& {McIntyre}, M.~E. 2011, Journal of Fluid Mechanics, 677, 445.
  \eprint{1005.5482}

\bibitem[{{Young}(2005)}]{YoungFIPquietsun2005A&A}
{Young}, P.~R. 2005, \aap, 439, 361. \eprint{arXiv:astro-ph/0503038}

\bibitem[{{Zweibel} \& {Gough}(1995)}]{egzdog1995ESASP}
{Zweibel}, E.~G., \& {Gough}, D.~O. 1995, in Helioseismology, vol. 376 of ESA
  Special Publication, 73

\end{thebibliography}

\end{document}